\documentclass[reprint, showpacs]{revtex4-1}
\usepackage{graphicx}
\usepackage{amsmath}
\usepackage{amsbsy}
\usepackage{natbib}
\usepackage{tikz}

\newcommand{\average}[1]{\left<#1\right>}

\newcommand{\hitr}{\mathbf{v}}
\newcommand{\hitru}{\mathbf{u}}
\newcommand{\lega}{\mathbf{r}}

\newcommand{\parcia}[2]{\frac{\partial\,#1}{\partial{#2}}}
\newcommand{\prej}[1]{#1_{\scriptscriptstyle{-}}}
\newcommand{\potem}[1]{#1_{\scriptscriptstyle{+}}}
\newcommand{\distpi}[1]{\rho_{\boldsymbol{\pi}}(#1)}
\DeclareMathOperator\erf{erf}

\begin{document}
\title{Exponential Fermi acceleration in general time-dependent billiards}
\author{Benjamin Batisti\'c}

\affiliation{CAMTP - Center for Applied Mathematics and Theoretical
Physics, University of Maribor, Krekova 2, SI-2000 Maribor, Slovenia}

\date{\today}


\begin{abstract}
  It is shown, that under very general conditions, a generic time-dependent billiard, for which a
  phase-space of corresponding static (frozen) billiards is of the mixed
  type, exhibits the exponential Fermi acceleration in the adiabatic limit.
  The velocity dynamics in the adiabatic regime is represented
  as an integral over a path through the abstract
  space of invariant components of corresponding static billiards, where
  the paths are generated probabilistically in terms of
  transition-probability matrices. We study the statistical properties of
  possible paths and deduce the conditions for the exponential Fermi acceleration.
  The exponential Fermi acceleration and theoretical concepts presented in
  the paper are demonstrated numerically in four different time-dependent
  billiards.
\end{abstract}

\maketitle

\section{Introduction}

An unbounded energy growth of particles in a time-dependent potential
is known as Fermi acceleration (FA), which was first proposed by Fermi
\cite{Fermi1949} to
explain the high energies of cosmic particles as a consequence of repeated 
collisions with moving interstellar magnetic domains.
Nowadays different models of FA are investigated in many 
areas of physics, such as astrophysics \cite{Kotera2011}, plasma physics \cite{Bian2013}, atom optics
\cite{Saif1998} and 
time-dependent billiards, which are the subject of this paper. 

Billiards are very simple and generic dynamical systems
of a fundamental importance for theoretical as well as numerical
investigations in classical \cite{Tabachnikov,Chernov,Koiller1995} and quantum mechanics
\cite{Batistic2013,Liss2013}. Billiards have been realized
also experimentally as microwave cavities \cite{stoe}, acoustic resonators,
optical laser resonators \cite{Gmachl1998} and quantum dots
\cite{Ponomarenko2008}. 
The first time-dependent billiard investigated in the context of FA was
the one dimensional Fermi-Ulam model (a particle between the moving walls)
\cite{Ulam1961}, for which it is nowadays known that it
does not permit FA if a motion of the walls is sufficiently smooth, due
to the existence of invariant tori which suppress the global energy
transport \cite{Lieberman1972}. The presence of chaos in two (or higher)
dimensional billiards make such an unbounded energy transport possible
\cite{Loskutov1999}.

Two dimensional periodic time-dependent billiards 
are the subject of intense investigations for almost two decades. 
Numerical studies suggest that asymptotically the average velocity
obeys the power law,
\begin{equation}
\average{v}\propto n^{\beta},
\label{eqPowLaw}
\end{equation}
with respect to the
number of collisions $n$, where several different values of the \textit{acceleration exponent} $\beta$
were observed
\cite{Lenz2008,Shah2011,Leonel2009a,Kamphorst2007,Carvalho2006b,Loskutov1999}.
The velocity dynamics is strongly related to the dynamical 
properties of a continuous set of \textit{corresponding static billiards}
which coincide with different shapes of a time-dependent billiard.
If all corresponding static billiards of a time-dependent billiard 
are ergodic then in general
$\beta=1/2$ \cite{Gelfreich2008a}, except
if the billiard motion is shape-preserving: in this
case $\beta$ depends only on the rotational properties of the billiard and
can have only one of the three possible values $\{0,1/6,1/4\}$
\cite{Batistic2011,Batistic2012,Batistic2014}.
However, if the dynamics is not ergodic then
$\beta$ could be greater than $1/2$ \cite{Leonel2009a}, moreover, it can even reach a
theoretical maximum asymptotic value $\beta=1$, which
corresponds to the exponential acceleration in the continuous time
\cite{Shah2011}.

Recently a lot of attention has been given to the possibility of a very
efficient unbounded exponential acceleration of particles in time-dependent
billiards.
It was shown theoretically, under very general conditions, 
that in time-dependent billiards possessing the
chaotic component, there exist trajectories of zero measure which
accelerate exponentially fast \cite{Gelfreich2008a}. 
However, under some circumstances the 
exponential acceleration can take place for most initial conditions.
This was first demonstrated in the rectangular billiard with the oscillating
bar \cite{Shah2010,Liebchen2011,Shah2013}, and then in a class of chaotic billiards
which undergo a separation of ergodic components by physically splitting
the billiard in several unconnected parts \cite{Gelfreich2011,Gelfreich2012}.
Recently it was shown in the study of an oscillating mushroom billiard,
how in this particular example the presence of the regular component
results in the exponential acceleration \cite{Gelfreich2013arxiv}.
Although the phenomenon of exponential acceleration is understood in
particular special examples, general insights have been lacking.

In this paper we consider the velocity dynamics in the adiabatic regime
and deduce general conditions for the exponential acceleration in time-dependent billiards.
The basic idea is to represent the motion of fast
variables as a Markov model of
a transport between the invariant components of corresponding static
billiards. 
It is shown that the exponential
acceleration arises if in the strict adiabatic limit, 
the number of possible paths through the space of
invariant components proliferate exponentially in time.
This condition is expected to be fulfilled if a corresponding static
billiards of a time-dependent billiard have more than one invariant
component, e.g. mixed type billiards.
We also clearly demonstrate the exponential acceleration numerically in
several different time-dependent billiards.

\section{Theory} \label{sectionTheroy}

The state of the particle in a time-dependent billiard is
described by the set $\{\lega,\theta,v,t\}$, where $\lega$ is a two-component position vector,
$\theta$ is a direction (angle) of a particle velocity vector
$\hitr=v\,(\cos{\theta},\sin{\theta})$,
$v=\|\hitr\|$ is a particle velocity and $t$ is time.
The motion of the particle is restricted to the billiard domain which is
periodically changing with time. 

The only force that acts on a particle in a billiard is that of a
boundary at collisions. Between collisions a particle velocity vector is preserved.
Collisions are elastic, which means that at
a collision, in a
reference frame in which the collision point is at rest, the normal
component of the velocity vector changes sign, while the
tangential component remains unchanged.

A velocity vector after the $n$-th collision at the collision point $\lega_n$ equals
\begin{equation}
  \hitr_{n}=\hitr_{n-1}-2\,P_{n}\,(\hitr_{n-1}-\hitru_{n}),
  \label{eqBarV}
\end{equation}
where $P$ is a projection matrix onto the normal to the boundary at
$\lega_n$, and $\hitru_n$ is the velocity vector of the boundary at $\lega_n$.
Squaring Eq. (\ref{eqBarV}) gives
\begin{equation}
  v_n^2=v_{n-1}^2 - 4\,\hitru_n\cdot P_{n}\,(\hitr_{n-1}-\hitru_{n}),
  \label{eqBarVsq}
\end{equation}
where we have used the properties of the projection matrix, $P^T=P$ and
$P^TP=P$. Now, using 
\begin{equation}
  v_n^2-v_{n-1}^2=\left( v_n-v_{n-1} \right)\left( v_n+v_{n-1} \right)
  \label{eqRel}
\end{equation}
and Eq. (\ref{eqBarV}) in Eq. (\ref{eqBarVsq}), we find
the change of the velocity at the $n$-th collision
\begin{equation}
  v_{n}-v_{n-1}=2\,
  \left(\frac{\hitr_{n}-\hitr_{n-1}}{v_{n} + v_{n-1}}\right)
  \cdot\hitru_{n}.
    \label{eqTest}
\end{equation}
If the velocity of the
boundary is zero, the particle velocity is preserved.

The objective of the paper is to understand under what conditions does the
sequence $v_n$ on average increase exponentially in time.
In the limit of large particle velocities,
the exponential acceleration in time corresponds to the
linear acceleration with respect to the number of collisions $n$,
which corresponds to the acceleration exponent $\beta=1$, as defined
in (\ref{eqPowLaw}).
This is because the number of collisions on the fixed time interval
is proportional to the particle
velocity,
\begin{equation}
  \Delta{n} \propto
  \frac{v}{\bar{\ell}}\,\Delta{t},
  \label{eqLinNtoExpT}
\end{equation}
where 
$\bar{\ell}$ is an average distance between two collisions on the time
interval $\Delta{t}$, which is large compared to a period of
billiard oscillations, but small enough to neglect the variations of $v$. 
It has to be assumed that in the adiabatic limit $\bar{\ell}$ is positive and independent of $v$, 
which is a reasonable assumption in general time-dependent billiards. 
For the extended discussion see \cite{Gelfreich2011}. We shall show in
numerical examples in next section that $\beta=1$ indeed corresponds to the exponential
acceleration.

In the adiabatic regime the velocity of the particle $v$ is much bigger
than any velocity of the boundary $u=\|\hitru\|$ and the time between two collisions is
much smaller than a period of a billiard motion. 
For a small but finite time interval $\delta{t}$ at time $t$,
on which the billiard changes
only very slightly, the following inequality is satisfied in the adiabatic
regime,
\begin{equation}
  u\,\delta{t}\ll \average{\ell}\ll v\,\delta{t},
  \label{eqAdiabaticIneq}
\end{equation}
where $\average{\ell}$ is an average distance between two collisions on a time
interval $\delta{t}$ at time $t$.
If (\ref{eqAdiabaticIneq}) is satisfied then a trajectory on the
time interval $\delta{t}$ around some time $t$ is approximately the same as if the particle
would be in the corresponding static billiard at time $t$, where
a corresponding static billiard at time $t$, is a static billiard
$(u\equiv0)$ which
has the same boundary as a time-dependent billiard at time $t$.
In the adiabatic limit, the geometry of trajectories in a time-dependent
billiard becomes independent of the particle velocity,
the same as in a static billiard.

In the adiabatic regime, the change of the particle velocity at a collision 
can be considered to depend only on $\left\{\lega,\theta,t\right\}$, as
for example
\begin{equation}
  v_{n}-v_{n-1}\approx
  \left(\frac{\hitr_{n}}{v_n}-\frac{\hitr_{n-1}}{v_{n-1}}\right) \cdot\hitru_{n},
    \label{eqTestAprox}
\end{equation}
which follows from (\ref{eqTest}) and the approximation
$v_{n}\approx v_{n-1}$.
Thus, formally, in the adiabatic regime, the particle velocity can be
approximately written as a path integral
\begin{equation}
  v(t_1)=v(t_0)+\int_{s(t_0)}^{s(t_1)}ds\,f(\lega(s),\theta(s),t(s)),
    \label{eqPathInteg}
\end{equation}
over a trajectory parametrized with $s$, 
where $s$ is a geometrical length of the path in the configuration space, and $f(\lega,\theta,t)$ is
a field independent of $v$. By parametrizing the trajectory 
in terms of time $t$ and using $ds=v\,dt$, the integral equation
(\ref{eqPathInteg}) can be equivalently written in a form of a differential equation
\begin{equation}
  \dot{v}=v\,f(\lega(t),\theta(t),t),
  \label{eqKompaktnoF}
\end{equation}
where dot denotes the time derivative.
A possible definition of the field $f(\lega,\theta,t)$ is presented in
Appendix A.
However, the actual definition of the field $f(\lega,\theta,t)$ is not important for
general conclusions of the theory that follows.

The integration of Eq. (\ref{eqKompaktnoF}) gives
\begin{equation}
  v(t)=v(t_0)\,e^{F(\mathcal{S})},
  \label{eqVnCyc}
\end{equation}
where we have introduced 
\begin{equation}
  F(\mathcal{S})=\int_{t_0}^{t} f(\lega(t),\theta(t),t)\,dt,
  \label{eqDefF}
\end{equation}
which is the integral of $f(\lega,\theta,t)$ along a trajectory $\mathcal{S}$ on the time
interval from $t_0$ to $t$.

Our goal is to describe the statistical properties of $F$ and deduce the
conditions for the exponential acceleration. 
We are going to
introduce a discrete time and represent the dynamics of the fast
variables $\left\{\lega,\theta\right\}$ as a stochastic hopping between the invariant components of
corresponding static billiards at discrete instances of time,
exploiting the fact that on a sufficiently small time interval
$\delta{t}$ and for a sufficiently big particle velocity $v$, the motion of
the fast variables is restricted to (and ergodic on) a single invariant
component of a corresponding static billiard.  

We divide the time interval of one period $T$ into $N$ small intervals of
length $\delta{t}=T/N$ on which the billiard can be considered static
and introduce a discrete time $j\in\{1,2\ldots\}$.
In the
adiabatic regime,
on a time interval $\delta{t}$ at time $j$ 
the motion of the fast variables $\left\{\lega,\theta\right\}$ 
is restricted to only one of the invariant components $\{\zeta^{j}_{n}\}$ 
of the corresponding static billiard at time $j$, 
where $n\in\left\{1,2\ldots\right\}$ index invariant components. 

We have assumed that there are in general countable many invariant components in a
static billiard. This is not exactly true because there is a continuum of
invariant tori in a regular domain if this is present in a billiard. Thus, for consistency,
we can consider a regular domain partitioned into a countable many
invariant components which are very thin layers of invariant tori.
On the other hand, we consider a connected chaotic domain as a single
invariant component, neglecting the zero measure set of isolated periodic
orbits.

In the adiabatic regime,
almost every trajectory on any invariant component $\zeta^j_n$ 
uniformly covers $\zeta^j_n$
within the time interval $\delta{t}$.
Thus the integral over $f$,
along almost any trajectory segment on the interval $\delta{t}$,
that lives in the invariant component $\zeta^{j}_n$ at time $j$,
approximately equals $\delta{F}\approx\delta{t}\,\bar{f}_{\zeta^{j}_{n}}$,
where $\bar{f}_\zeta$ denotes the average of $f$ on the invariant component
$\zeta$. 

We shall call a chronologically ordered sequence
of invariant components $\{\zeta^{j}_{n_j}\}$ a $\zeta$-trajectory.
In the adiabatic regime,
every trajectory can be represented as a $\zeta$-trajectory. 
We shall not distinguish between trajectories which are represented with
the same $\zeta$-trajectory. In other words, the $\zeta$-trajectory
represents a maximal resolution of our theory.
In terms of a $\zeta$-trajectory, $F$ is an "integral" over the path
through the space of invariant components of corresponding static
billiards,
\begin{equation}
  F\approx\sum_{j}\delta{t}\,\bar{f}_{\zeta^{j}_{n_j}}.
  \label{eqProbaF}
\end{equation}

$\zeta$-trajectories are generated probabilistically 
in terms of transition matrices $\{P^j\}$, 
where a matrix element $P^{j}_{n,m}$ is a probability for the transition 
$\zeta^j_m\rightarrow\zeta^{j+1}_n$ between two invariant components of two
successive corresponding static billiards at times $j$ and $j+1$, respectively.
A transition-probability $P^{j}_{n,m}$ is bounded between 0 and 1, and
can be only a monotonic function of the particle velocity $v$,
thus, in the adiabatic limit, it either vanishes or it converges to a
positive constant independent of $v$. In the adiabatic regime, we
consider $\{P^j\}$ to be constant matrices independent of $v$. 

If at least some of transition matrices $\left\{ P^j \right\}$ are stochastic
matrices, which means that at lest
some matrix elements are different than 0 or 1, then a number of possible
$\zeta$-trajectories increases exponentially with increasing $j$. 

A transition matrix $M=P^{N}\ldots P^{2}\,P^{1}$ determines
transition probabilities between invariant components of an initial
corresponding static billiard after one cycle of a billiard motion. 
If not all invariant components of an initial corresponding static billiard
are connected then the transition matrix $M$ is a block matrix. In this
case we can consider each block separately as an independent system.
In the following, let the matrix $M$ correspond to a single block which
represents a subset of invariant
components which are connected. Then, by the 
Perron-Frobenius theorem, there exists a unique invariant probability vector
${\boldsymbol{\pi}}$, such that ${\boldsymbol{\pi}}=M{\boldsymbol{\pi}}$,
and the sequence of the powers of $M$ converges to a stationary matrix
$M^{\infty}$ which has all columns equal to $\boldsymbol{\pi}$. 
A vector $\boldsymbol{\pi}$ is an invariant discrete probability
distribution on a discrete set of invariant components of an initial
corresponding static billiard.

Let $F_m$ denote a value of $F$ after $m$ cycles of
a billiard motion and let  $\rho_{\boldsymbol{\pi}}(F_m)$ 
be a probability distribution for
$F_m$ with respect to an invariant probability distribution 
$\boldsymbol{\pi}$. All possible values of $F_m$ are determined
by the transition matrix $M$ and
by a pool of all possible values of $F_1$ corresponding to all possible
$\zeta$-trajectories within one cycle of a billiard motion. 
Thus, for a particular billiard, it is essential to understand what are
possible values of $F_1$. We shall show, using the Liouville theorem, that if for some
$\zeta$-trajectories the corresponding values of $F_1$ do not vanish in the adiabatic limit, 
then this leads to the exponential acceleration on average, which is
nontrivial, since $F_1$ can be negative as well, resulting in the
exponential deceleration according to Eq. (\ref{eqVnCyc}).

There are three types of time-dependent billiards in which $F_1$
vanishes in the adiabatic limit for almost all initial conditions and
consequently the acceleration is slower than exponential:
\begin{enumerate}
  \item 
A time-dependent billiard in which 
all corresponding static billiards have only one invariant component
which is necessarily ergodic, excluding a zero measure set of
isolated periodic orbits.
In this case there is only one $\zeta$-trajectory, for which
$F_1\rightarrow0$
according to the adiabatic law
$v_1\,\sqrt{\mathcal{A}_1}=v_0\,\sqrt{\mathcal{A}_0}$, where $\mathcal{A}$
is an area of a billiard \cite{Hertz1910}.
In the adiabatic regime, the fluctuations of $F_1$, denoted by
$\delta{F_1}$, scale with the velocity as $\delta{F_1}\propto
1/\sqrt{v_0}$, which follows from the following consideration.
The difference $\delta{v}=v_1-v_0$ between the initial velocity $v_0$ and the final velocity
$v_1$ is a sum of $n\propto v_0$ terms from each collision 
within one cycle of the billiard motion. If these terms are uncorrelated
then $\average{\delta{v}^2}\approx \kappa^2\,n$ and
\begin{equation}
  \average{\delta{F_1}^2}=\average{\left(\log{\frac{v_0+\delta{v}}{v_0}}\right)^2}
  \approx \frac{\kappa^2\,n}{v_0^2}=\frac{\kappa^2}{v_0}
    \label{eqFest}
\end{equation}
where $\kappa^2$ is some number independent of the velocity. 
So, the distribution $\rho_{\boldsymbol{\pi}}(F_1)$ depends on the
velocity of the initial
ensemble. We expect that $\rho_{\boldsymbol{\pi}}(F_1)$ satisfies the following
scaling property
\begin{equation}
  \frac{1}{\sqrt{v_a}}\,\rho_{\boldsymbol{\pi}}(\sqrt{v_a}\,F_1)=
  \frac{1}{\sqrt{v_b}}\,\rho_{\boldsymbol{\pi}}(\sqrt{v_b}\,F_1)
  \label{eqDistChaosF1}
\end{equation}
for every pair of sufficiently large initial velocities $v_a$ and $v_b$ of an ensemble of
particles.

  \item
A time-dependent billiard in which all corresponding static billiards
are integrable \cite{Lenz2010}. 
In this case the adiabatic invariance of actions \cite{Arnold} ensures
that $F_1\rightarrow0$ for every
$\zeta$-trajectory. Moreover, a matrix $M$ is an identity matrix for
every initial corresponding static billiard, thus a number of
$\zeta$-trajectories is constant, equal to a number of
invariant components. In the case of a time dependent ellipse, as
demonstrated in \cite{Lenz2009,Lenz2010}, the motion of the boundary
induces a chaotic layer around the separatrices of corresponding static
billiards, which plays a crucial role
in the acceleration. A width of the chaotic layer depends
on the particle velocity and vanishes in the strict limit
$v\rightarrow\infty$. The chaotic layer 
can not be considered as an invariant component of a
corresponding static billiard.
  \item
A billiard which undergoes shape-preserving
transformations \cite{Batistic2014}, such that a distance $\ell$ between each pair of points
on a boundary changes by the same proportion, which means that
$\dot{\ell}/\ell$ is constant, where $\dot{\ell}$ is a time derivative of
$\ell$. If a billiard driving is periodic this implies $F_1\rightarrow0$ as shown in the
Appendix B. 
In the adiabatic limit, 
in a reference frame in which a shape preserving billiard
is at rest, the particle dynamics quickly converges
to the dynamics of a static billiard \cite{Batistic2014}.
This implies that in the adiabatic limit the
transport between invariant components of a static billiard is suppressed
and consequently a number of $\zeta$-trajectories is
constant, equal to a number of invariant components.
\end{enumerate}

If some corresponding static billiards of a time-dependent billiard host more
than one invariant component then in such a time-dependent billiard 
there are many possible $\zeta$-trajectories. If transitions
between invariant components are stochastic, then
$\zeta$-trajectories are mixing in a sense that they can 
cross on a single invariant component. In this case a number of possible
$\zeta$-trajectories increases exponentially in time.

\begin{figure}
  \vspace{0.2cm}
\includegraphics[width=3.8cm]{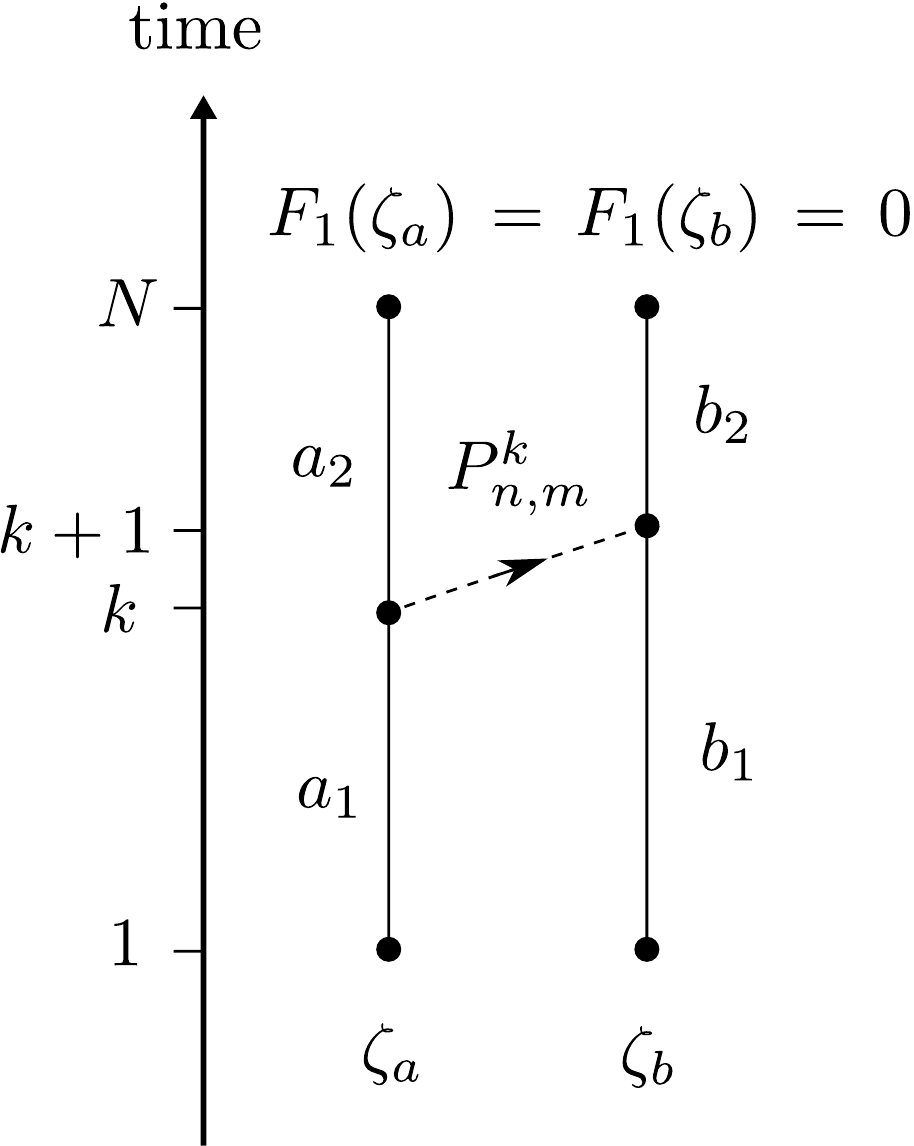}
\caption{Two $\zeta$-trajectories, namely $\zeta_a$ and
    $\zeta_b$, which are represented as straight vertical lines, for which
    $F_1(\zeta_a)=a_1+a_2=0$ and $F_1(\zeta_b)=b_1+b_2=0$, are mixing
    with the probability $P^k_{n,m}$ from the invariant
    component $n$ which is the state on the trajectory $\zeta_a$ at time
    $j=k$ to the invariant component $m$ which is the state of the
    trajectory $\zeta_b$ at time $j=k+1$. This allows an additional
$\zeta$-trajectory for which $F_1=a_1+b_2\neq0$ if $a_1\neq b_1$.}
\label{figTrans}
\end{figure}

In a generic case the mixing of $\zeta$-trajectories implies the existence
of $\zeta$-trajectories for which $F_1\neq0$. This can be proved by
contradiction. Assume that for all $\zeta$-trajectories $F_1=0$, then,
because some $\zeta$-trajectories are mixing, a pair of mixing 
$\zeta$-trajectories can be found for which $F_1=0$. Let $\zeta_a$ and
$\zeta_b$ be two such $\zeta$-trajectories. See Fig. \ref{figTrans}. 
Let the mixing be such that at
time $k$ there is a nonzero probability $P^k_{n,m}$ to switch from
$\zeta_a$ to $\zeta_b$. Thus, there exists a $\zeta$-trajectory, denoted by
$\zeta_{ab}$, which follows $\zeta_a$ from time $j=1$ up to time $j=k$ and
then switches with the probability $P^k_{n,m}$ to $\zeta_b$ at time
$j=k+1$ and stick to it until a cycle of a billiard motion completes.
Let $a_1$ and $b_1$ denote the changes of $F$ on the time interval from
$j=1$ to $j=k$ and let $a_2$ and $b_2$ denote the changes of $F$ on the
time interval from $j=k+1$ to $j=N$ on the trajectories $\zeta_a$ and $\zeta_b$,
respectively. By the assumption
\begin{eqnarray}
    F_1(\zeta_a)&=& a_1+a_2=0,\nonumber\\
    F_1(\zeta_b)&=& b_1+b_2=0,
    \label{eqFzetAinB}
\end{eqnarray}
and thus
\begin{equation}
    F_1(\zeta_{ab})=a_1+b_2.
    \label{eqFzetab}
\end{equation}
If $a_1=b_1$ and $a_2=b_2$ then $F_1(\zeta_{ab})=0$ as
well. 
However, we expect that in a generic case different
$\zeta$-trajectories are not correlated, which means that in general $a_1\neq b_1$ and
$a_2\neq b_2$, and then, according to Eq. (\ref{eqFzetab})$, F_1(\zeta_{ab})\neq0$.
Thus, we have found a $\zeta$-trajectory for which $F_1\neq0$, which
contradicts the original assumption that for all $\zeta$-trajectories
$F_1=0$.

Now, knowing that that for some $\zeta$-trajectories $F_1$ is nonvanishing, 
we shall show that this must result in the exponential
acceleration, which is nontrivial since $F_1$ can be negative as well.

Let $\gamma$ be a finite number of cycles of a billiard motion 
after which $M^{\gamma}$ ($M$ raised to the power of $\gamma$) can be
considered sufficiently close to $M^{\infty}$, which is effectively a
number of cycles after which correlations between initial and
final states of $\zeta$-trajectories are lost. 
Note, that if at least one corresponding static billiard is ergodic
(has only one invariant component) then $\gamma=1$.
By the definition of $\gamma$, 
a probability distribution $\rho_{\boldsymbol{\pi}}(F_{\gamma\,k})$ for
$F$ after $m=\gamma\,k$ cycles of a billiard motion,
where $k$ is some positive integer, equals the
$k$-fold convolution power of $\rho_{\boldsymbol{\pi}}(F_\gamma)$.
Using this fact and the distribution of the velocity in
terms of $F$,
\begin{equation}
  \rho(v)=\int dF\,\delta(v-v_0\,e^F)\,\rho_{\boldsymbol{\pi}}(F),
  \label{eqFtoVtransf}
\end{equation}
we find that a corresponding average velocity after $m=\gamma\,k$ cycles equals
\begin{equation}
  \average{v_{\gamma\,k}}= v_0\,\langle e^{F_\gamma}\rangle^k,
  \label{eqAkcija}
\end{equation}
where $v_0$ is an initial velocity.

Now we show that the incompressibility of the phase-space flow (Liouville
theorem)
implies $\average{e^{F_\gamma}}>1$ and thus the exponential acceleration.
For the arguments sake,
suppose $\gamma$ is big enough for $\rho_{\boldsymbol{\pi}}(F_\gamma)$ to
be approximately Gaussian with the mean $\mu$ and the width $\sigma>0$,
\begin{equation}
  \rho_{\boldsymbol{\pi}}(F_\gamma)=\frac{1}{2\,\pi\,\sigma^2}\,e^{-\frac{(F_\gamma-\mu)^2}{2\,\sigma^2}}.
  \label{eqFgamaGaus}
\end{equation}
Consider some finite velocity $v_c$ and denote with 
$\Omega_{c}$ the volume of the phase-space below $v_c$.
Take some large part of the phase-space above $v_c$ having
the volume $\Omega\gg\Omega_{c}$ and the initial velocity distribution
$\rho(v_0)$.
A phase-space volume $\Omega_{v<v_c}$ that leaks below $v_c$ after
$m=\gamma\,k$ cycles equals
\begin{eqnarray}
  \Omega_{v<v_c}&=&\Omega\int_0^{v_c} dv\,\rho(v)\nonumber\\
  &=& 
  \frac{\Omega}{2}\int
  \left[1-\erf\left(\frac{\mu\,k+\log(v_0/v_c)}{\sqrt{2\,k\,\sigma^2}}\right)\right]
  \rho(v_0)\,dv_0,\nonumber\\
  \label{eqGausProb}
\end{eqnarray}
where we have used the fact that the phase-space volume is proportional to the probability.
From (\ref{eqGausProb}) we see that if $\mu<0$ or $\mu=0$ then in the limit $k\rightarrow\infty$ 
the phase-space volume $\Omega_{v<v_c}$ converges to
$\Omega$ or $\Omega/2$, respectively. But the amount of the phase-space
volume that can be occupied below $v_c$ is limited, $\Omega_{v<v_c}\leq\Omega_c$,
thus 
$\Omega_{v<v_c}\rightarrow\Omega$ or $\Omega_{v<v_c}\rightarrow\Omega/2$
contradicts either the initial assumption $\Omega\gg\Omega_c$ or the Liouville
theorem. Therefore, if $\sigma>0$ then $\mu>0$, which implies
\begin{equation}
  \average{e^{F_\gamma}}=
  \int^{\infty}_{-\infty}
  \frac{dF_{\gamma}}{2\,\pi\,\sigma^2}\,e^{-\frac{(F_{\gamma}-\mu)^2}{2\,\sigma^2}}=
  e^{\mu+\sigma^2/2}>1
  \label{eqGaussModel}
\end{equation}
and thus Eq. (\ref{eqAkcija}) implies the exponential acceleration.
This is the central result of the paper.

\section{Numerical results}

The exponential acceleration was already demonstrated in time-dependent
billiards for which a number of physically connected parts of a
billiard domain vary with time \cite{Gelfreich2012}. The exponential acceleration
in such a time-dependent billiard is easily explained with the theory
we have developed in section \ref{sectionTheroy}. Recently the
exponential acceleration was also demonstrated in a time-dependent
mushroom billiard, which is a nonsmooth billiard with sharply separated
regular and chaotic domains \cite{Gelfreich2013arxiv}. However,
there has been no clear demonstration of the exponential acceleration in a
smooth time-dependent billiard of the mixed type.

Demonstrating the exponential acceleration numerically could be a demanding
task \cite{Shah2011}.
The problem is, that only a finite
number of collisions can be computed, which might not be enough to
demonstrate exponential acceleration in an affordable amount of time.
The exponential acceleration can not be demonstrated numerically if the
asymptotic regime occurs at so high velocities 
that we can not afford to compute all the
collisions of a reasonable big ensemble within one cycle of the billiard
motion and show at least that $\rho_{\boldsymbol{\pi}}(F_1)$ is
asymptotically independent of the particle velocity.

The asymptotic regime of the exponential acceleration 
arises when the distribution $\distpi{F_1}$
becomes effectively independent of the particle velocity.
A regime below the asymptotic regime is called a transient regime.

Consider a time-dependent billiard which is predominantly chaotic with
relatively small regular domains. 
In a transient regime the presence of regular domains in the phase-space
of corresponding static billiards is negligible and the billiard behaves
as a fully chaotic.
So, if $\sigma$
is a width of $\distpi{F_1}$, then, according to
Eq. (\ref{eqFest}), the asymptotic regime is reached when the particle velocity $v$
satisfies
\begin{equation}
  \sigma\gg \frac{\kappa}{\sqrt{v}}.
  \label{eqAsymCrit}
\end{equation}
The mechanism of exponential
acceleration prevails when the above condition is satisfied.

Long transient regimes are expected also if deformations of a
billiard shape are small such that a structure of the phase-space of
corresponding static billiards varies very little. In this case a
transport between regular and chaotic domains is weak, which renders the
mechanism of exponential acceleration weak as well.

In a transient regime we can observe a more efficient acceleration
than $\average{v}\propto n^{1/2}$, which is expected in
fully chaotic billiards. This can be also due to the presence of sticky
objects in a chaotic domain which act as quasi-invariant components for
sufficiently small but large particle velocities.

In the following we present numerical analysis of  four different time-dependent
billiards. Two of them were already studied before: Oval billiard
\cite{Leonel2009a} and Annular billiard \cite{Carvalho2006b,Carvalho2011}.
Although these billiards are of the mixed type the authors did not observe
the exponential acceleration because they did not reach the asymptotic
regime in their numerical simulations. We consider this two
billiards again and show that they indeed exhibit the
exponential acceleration. We have adjusted the parameters of the billiards in
order to make the mechanism of exponential acceleration stronger and the
transient regimes shorter.

We have also considered two time-dependent billiards which have not been studied
before: a time-dependent
Robnik billiard and a billiard which is a non-convex deformation of the
elliptical billiard. The last billiard is studied in depth and is
presented here as a main example.

\subsection{Main example}

In this subsection we consider a time-dependent billiard with 
a boundary which satisfies a time-dependent implicit equation
\begin{equation}
  x^2 + \frac{2\,y^2}{1 + a\,(1+\cos{t})\,\left( x^2-1 \right)}=1,
\label{eqTransfB}
\end{equation}
where $a$ is the deformation parameter.
The period of the billiard motion is $2\,\pi$.
The motion of the billiard alternates between expanding and contracting
phase in which the billiard passes over the same sequence of
corresponding static billiards, but in the reversed order.

\begin{figure}
  \vspace{0.2cm}
\includegraphics[width=8.2cm]{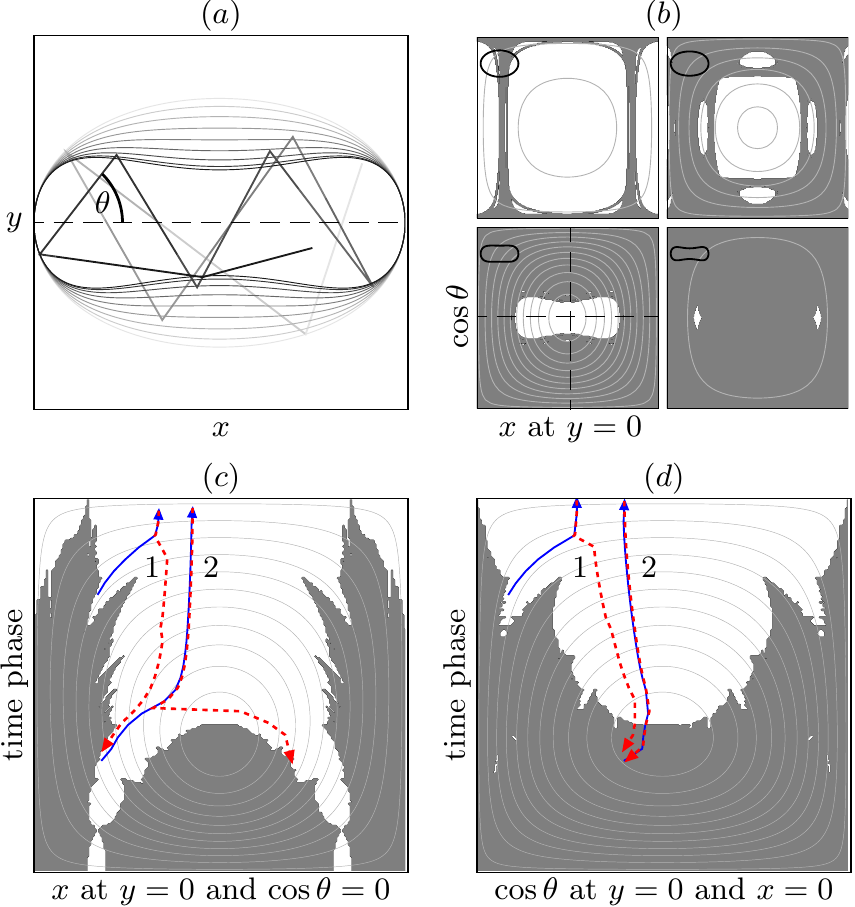}
\caption{
Numerical study of the phase-space of a time-dependent billiard defined in
(\ref{eqTransfB}) with the deformation parameter $a=0.3$ (a).
(b), (c), (d): 
Projections on the Poincare line of section $y=0$ (the dashed line in (a));
gray denotes chaotic and white regular regions of the
corresponding static billiards;
in light grey, contours of constant $\left|f'\right|$, Eq.
(\ref{eqDefAccFact}), on 11 equidistant levels between 0 and $f'_{\max}$.
(b): 
Phase space structures of four corresponding static billiards (up-left).
(c), (d): 
Time evolution of two slices (dashed lines in (b)). Expanding and
contracting phases are symmetric:
expanding phase = direction up, contracting phase = direction down.
Lines with arrows are fractions of two trajectories in two
different projections:
a time of one period was divided into 200 subintervals on which the value of
local minimum  of $x$ (and $\cos\theta$) of a trajectory was 
determined and used in the plot instead of all intersections with the
surface of section. 
Parts of trajectories in the chaotic region are not plotted. 
Solid blue and dashed red represent the expanding and contracting
phase, respectively.
The velocities of considered trajectories are $\sim10^5$.
Both trajectories start in the chaotic component at the beginning of the
expanding phase, which is at the bottom of the diagrams. The trajectory
1 is a typical example for which $F_1>0$ as can be seen from the
path trough the contours of constant $\left|f'\right|$, while the
trajectory 2 is symmetric and thus $F_1\rightarrow0$.
}
\end{figure}

\begin{figure}
  \vspace{0.2cm}
\includegraphics[width=8.2cm]{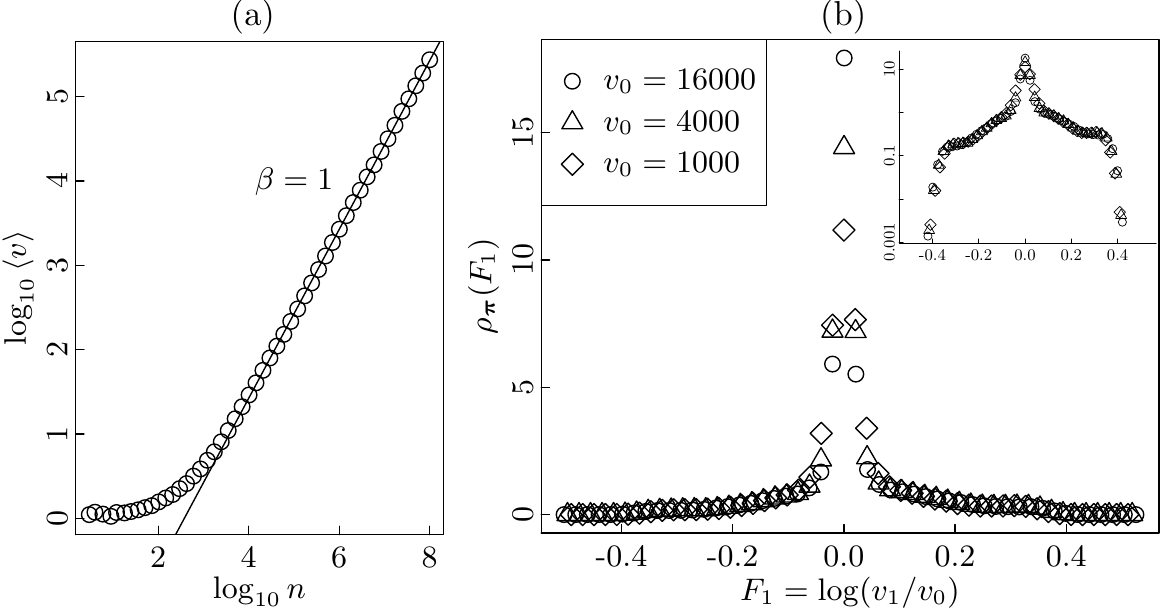}
\caption{
  (a): A linear increase of the average velocity with respect to the number of
  collisions $n$ (exponential acceleration in the continuous time $t$);
  $10^3$ initial conditions used.
  (b): The distribution of $F_1$ for different $v_0$
   and for
  $10^6$ initial conditions uniformly distributed in
  $\left\{\lega,\theta\right\}$ at $t=0$, when the billiard is almost
  completely chaotic (logarithmic scale in the subfigure).
  The central peak is growing as $\sqrt{v_0}$ and converges to the Dirac
  delta distribution. 
  The peak corresponds to symmetric $\zeta$-trajectories which pass over
  the same sequence of invariant components in both expanding and
  contracting phase. 
  The distribution $\rho_{{\boldsymbol{\pi}}}(F_1)$ is effectively independent of the
  velocity, has a positive mean and a finite width, which
  implies the exponential acceleration (a).
}
\end{figure}

\begin{figure}
  \vspace{0.2cm}
\includegraphics[width=5cm]{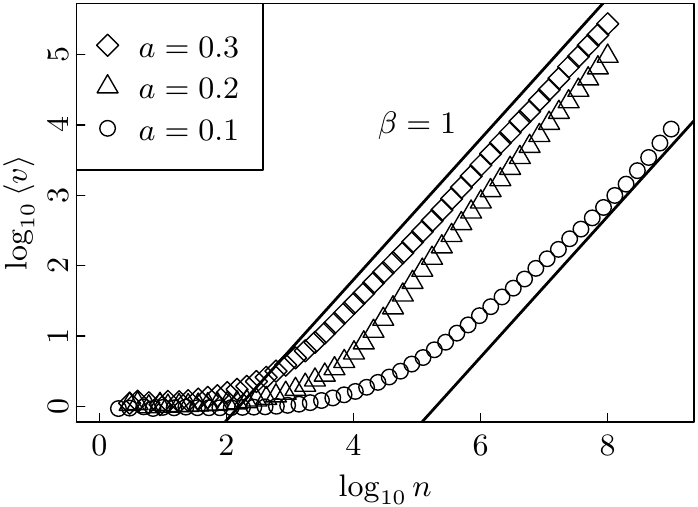}
\caption{
  The acceleration for different deformation parameters $a$ in a billiard
  with the boundary as defined in (\ref{eqTransfB}). We see that smaller
  deformations (smaller $a$) correspond to a slower transition to the
  exponential acceleration (solid lines with slopes 1).
}
\end{figure}

In the following
we shall closely consider the case $a=0.3$, shown schematically on
Fig. 1(a).
In this case,
the corresponding static billiards change from the
almost completely chaotic at $t=0$ to the completely regular (ellipse) at
the half period $t=\pi$. In between the structure of the phase-space is
mixed and the invariant phase-space structures are rapidly changing with
time, thus a very clear exponential acceleration is expected to be
observed.

In Figs. 1(b)-(d) different projections of the phase-space are presented,
in which a chaotic domain is colored gray and a regular domain is colored
white.
Together with the structures of the phase-space, 
we plot the contours of
constant $|f'|$, where $f'$ is defined in Appendix B in Eq.
(\ref{eqDefAccFact}) and is just one of the possible approximations of the
field $f(\lega,\theta,t)$ introduced in Eq. (\ref{eqPathInteg}). 
The contours help to demonstrate that the
acceleration of the particle, Eq.
(\ref{eqKompaktnoF}), is different in
different parts of the phase-space and that the average of $f$ is
different on different invariant components of the corresponding static
billiards. Thus, different $\zeta$-trajectories have different associated
values of $F_1$, where not all of them can be zero. 
Therefore the distribution of
$F_1$ must have a finite variance, as shown in Fig. 2(b), which according to the theory implies the
exponential acceleration, as shown in Fig. 2(a).

The theory predicts that $\rho_{\boldsymbol{\pi}}(F_1)$ has asymptotically a nonvanishing
width and a velocity independent shape.
As shown in Fig. 2(b) the
velocity dependence of $\rho_{\boldsymbol{\pi}}(F_1)$ is already barely visible
for velocities $v>10^3$, except for the central peak. This peak is a
consequence of the symmetry of the driving and converges to the Dirac
delta distribution in the limit $v\rightarrow\infty$.
As already mentioned, the motion of the billiard alternates between the expanding and the contracting
phases in which the billiard passes over the same sequence of
corresponding static billiards, but in the reversed order.
This symmetry implies that $F_1\rightarrow0$ for a symmetric $\zeta$-trajectory which passes over
the same sequence of invariant components in both expanding and
contracting phase. 
Let us decompose the distribution $\rho_{\boldsymbol{\pi}}(F_1)$ into
a sum 
\begin{equation}
    \rho_{\boldsymbol{\pi}}(F_1)=p_a\,\rho_a(F_1) + p_b\,\rho_b(F_1)
    \label{eqDecomp}
\end{equation}
where $\rho_b(F_1)$ is the distribution of $F_1$ for symmetric $\zeta$-trajectories,
and corresponds to the peak at $F_1=0$, in Fig. 2(b).
According to Eq. (\ref{eqDistChaosF1}), the width of the distribution $\rho_b(F_1)$
scales as $1/\sqrt{v_0}$ and its
height scales as $\sqrt{v_0}$.
Now, since the fraction of
symmetric $\zeta$-trajectories is constant, and thus $p_b$ is constant, 
the height of the central peak of
$\rho_{\boldsymbol{\pi}}(F_1)$ should scale as $\sqrt{v_0}$, in
agreement with the numerical results.

In Fig. 3 we show how the length of the transient regime depends on the deformation parameter
$a$, Eq. (\ref{eqTransfB}). If deformations of the billiard are small ($a=0.1$) then the structure of the
phase-space of corresponding static billiards does not change a lot.
This results in a reduced transport between invariant domains and
consequently in the weak mechanism of the exponential acceleration which
reveals itself only after a long transient regime.

\subsection{Oval billiard}

In this subsection we shall consider a time-dependent oval billiard which was
already studied in \cite{Leonel2009a}. The shape of the billiard 
is define in a polar coordinates as
\begin{equation}
  R(\theta)=(1+\eta_1)+\epsilon\,(1+\eta_2\,\cos{t})\,\cos{2\,\theta},
  \label{eqOvalB}
\end{equation}
where $\eta_1$ and $\eta_2$ are deformation parameters.
For convenience we have chosen the same symbols for the deformation
parameters as in \cite{Leonel2009a}.
In \cite{Leonel2009a} only small deformations of a boundary were
considered ($\eta_1\leq 0.1$ and $\eta_2\leq 0.1$) at $\epsilon=0.4$ for
which the whole phase-space is almost entirely chaotic, except for the
relatively small islands of regular motion. While the phase-space is
of the mixed type, though predominantly chaotic, the exponential
acceleration is still expected in the deep adiabatic limit. In
\cite{Leonel2009a}, however, the
regime of exponential acceleration was not jet reached and thus not
observed, although the numerical simulations had been evolved up to $10^9$
collisions and the velocities of the order of $10^3$ were reached.

\begin{figure}
  \vspace{0.2cm}
\includegraphics[width=8.3cm]{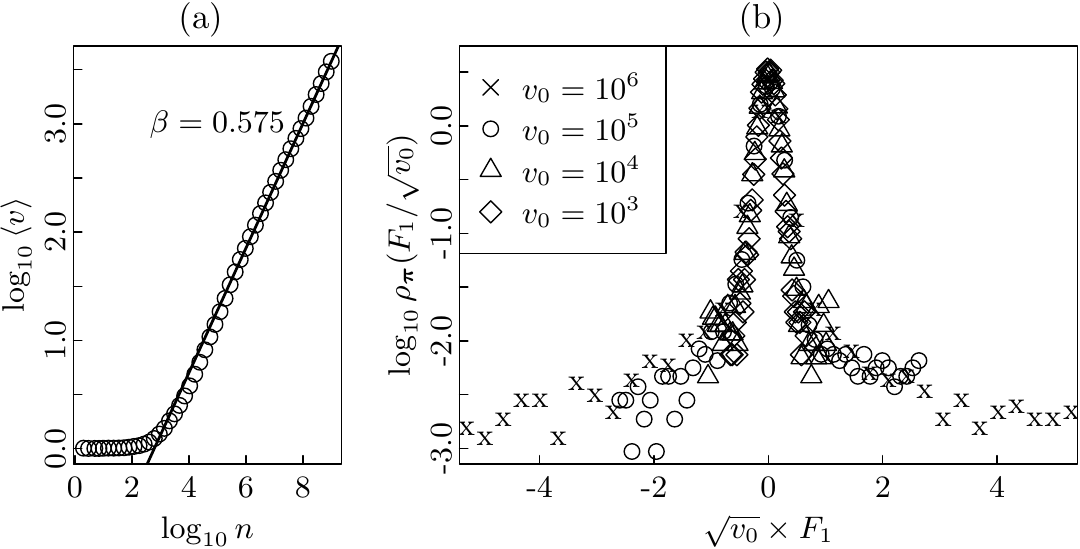}
\caption{
  The acceleration of particles ($10^3$ initial conditions at $v_0=1$) in time-dependent oval billiard with
  $\epsilon=0.4$ and deformation parameters $\eta_1=0$ and $\eta_2=0.1$
  (a). The acceleration exponent $\beta$ is far from the expected
  asymptotic value $\beta=1$. The
  distribution of $F_1$ steel strongly depends on the velocity (b), satisfying
  the scaling property (\ref{eqDistChaosF1}) for fully chaotic systems.
  However, we see that tails are already independent of the velocity,
  thus in the deep adiabatic limit the mechanism of exponential
  acceleration should prevail. The distributions in (b) are calculated
  with ensembles of $10^4$ initial conditions.
  \label{figOval0401}
}
\end{figure}

We have considered again the case $\epsilon=0.4$, $\eta_1=0$ and
$\eta_2=0.1$, which was considered already in
\cite{Leonel2009a}. As shown on Fig. \ref{figOval0401}(a), the exponential
acceleration is not observed even after $10^9$ collisions, where the local
acceleration exponent is $\beta\approx 0.575$. This is consistent with the
distribution of $F_1$, Fig. \ref{figOval0401}(b), where we can see that in
the range of velocities $10^3-10^6$ the distribution of $F_1$ effectively
satisfies the scaling relation (\ref{eqDistChaosF1}), which is valid for fully
chaotic systems. However, the tails of the distribution are
independent of the velocity, as expected, but in the concrete velocity range they are
too weak to dominate the acceleration. 

\begin{figure}
  \vspace{0.2cm}
\includegraphics[width=8.3cm]{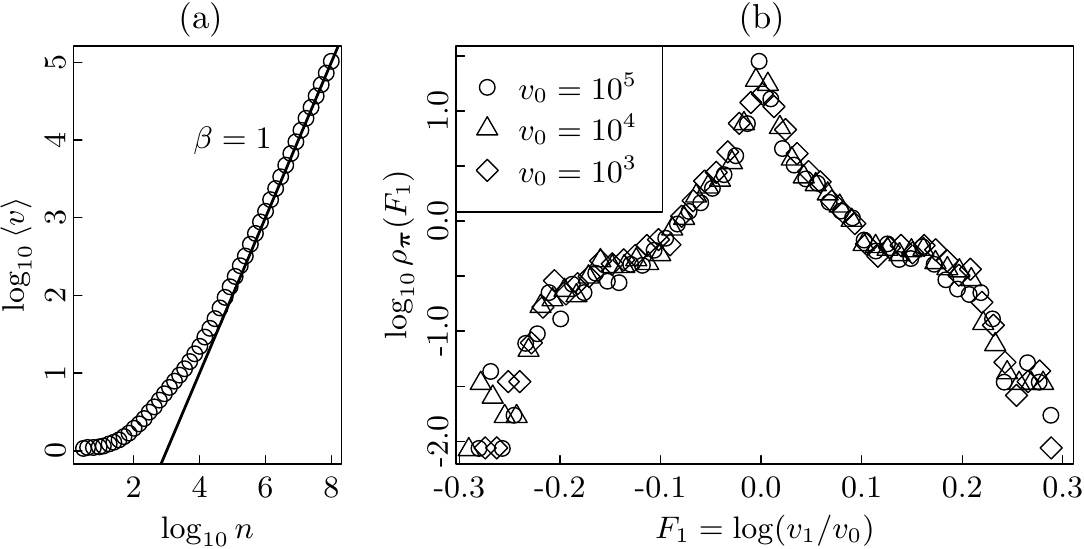}
\caption{
  The acceleration of particles ($10^3$ initial conditions at $v_0=1$) in time-dependent oval billiard with
  $\epsilon=0.2$ and deformation parameters $\eta_1=0$ and $\eta_2=0.9$
  (a). The acceleration exponent $\beta$ is approaching $\beta=1$. The
  distribution of $F_1$ is (effectively) independent of the velocity in
  the regime $v>10^3$ (b), which is the indicator of the exponential
  acceleration ($\beta=1$). The distributions in (b) are calculated
  with ensembles of $10^4$ initial conditions.
  \label{figOval0901}
}
\end{figure}

In order to demonstrate the exponential acceleration in a time-dependent
oval billiard we chose a set of parameters $\epsilon=0.2$, $\eta_1=0$ and
$\eta_2=0.9$, for which the phase-space of corresponding static billiards 
is more diverse and rapidly changes with time. This set of parameters was
not studied in \cite{Leonel2009a}. As expected, we see on Fig.
\ref{figOval0901} that this billiard exhibits a clear exponential
acceleration.

It is worth mentioning that in \cite{Leonel2009a} the authors studied the
case $\epsilon=0.4$ and $\eta_1=\eta_2=0.1$, which is a time dependent
scaling transformation of the billiard. As already mentioned in this case
the exponential acceleration is not possible and the acceleration exponent
equals $\beta=1/6$, which is theoretically explained in
\cite{Batistic2014}.

\subsection{Robnik billiard}

\begin{figure}
  \vspace{0.2cm}
\includegraphics[width=3.5cm]{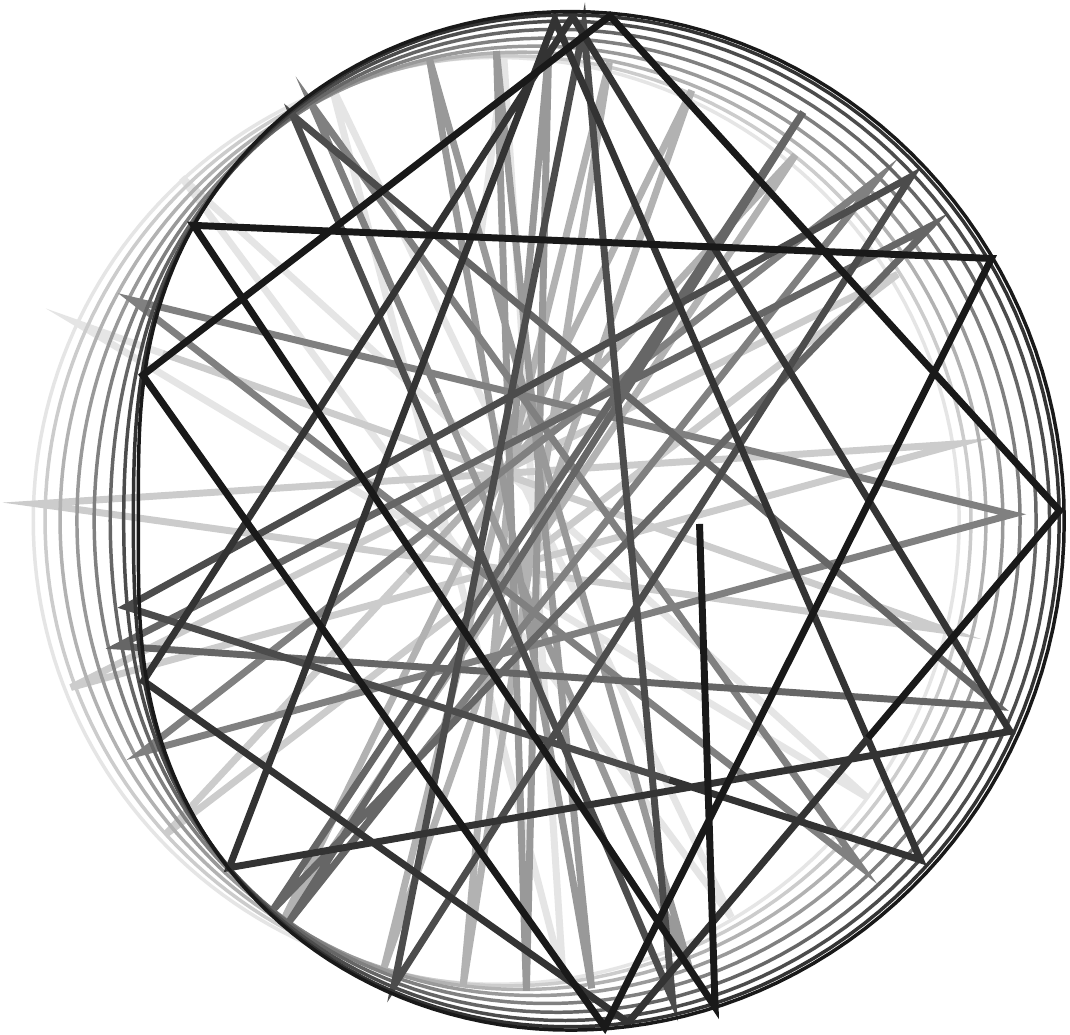}
\caption{
  A trajectory in the time dependent Robnik billiard (\ref{eqRobnikDef}).
  \label{figRobBil}
}
\end{figure}

In this subsection we consider a billiard with the boundary given in a
parametric form
\begin{eqnarray}
  x(s) &=& \cos(s) + \lambda(t)\,\cos(2\,s),\nonumber\\
  y(s) &=& \sin(s) + \lambda(t)\,\sin(2\,s)
  \label{eqRobnikDef}
\end{eqnarray}
where $s$ is a parameter that  runs from $0$ to $2\,\pi$, and where
\begin{equation}
  \lambda(t)=\frac{1-\cos{t}}{8}
  \label{eqRobnLamT}
\end{equation}
is the time-dependent deformation parameter, Fig. \ref{figRobBil}.

The set of static billiards (\ref{eqRobnikDef}) for fixed values of
$\lambda$ are known as Robnik billiards, introduced by Robnik \cite{Robnik1983}. 
For $\lambda=0$ at $t=0$, the billiard
boundary is a circle and the corresponding static billiard is
integrable. With the increasing $\lambda$ the boundary deforms and the phase
space of corresponding static billiards becomes mixed with the increasing
chaotic component. Finally for $\lambda=1/4$ at $t=\pi$ the billiard is
almost ergodic \cite{Hayli1987}.

\begin{figure}
  \vspace{0.2cm}
\includegraphics[width=8.3cm]{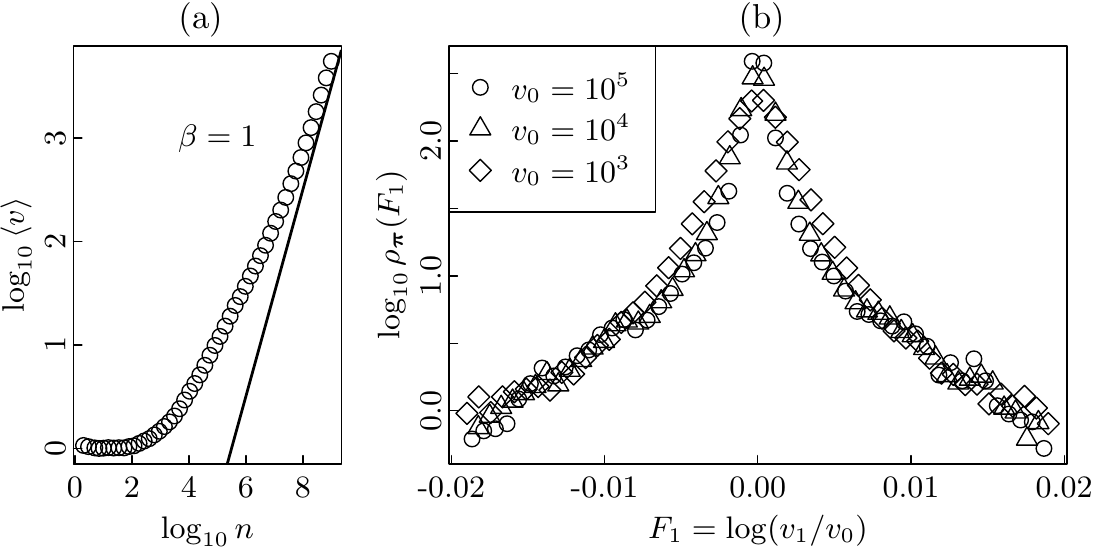}
\caption{
  The acceleration of particles ($10^3$ initial conditions at $v_0=1$) in time-dependent 
  Robnik billiard (\ref{eqRobnikDef}). The
  distribution of $F_1$ is (effectively) independent of the velocity in
  the regime $v>10^4$, while at $v=10^3$ the distribution is still
  noticeably wider (b). The exponential acceleration is clearly visible
  only after $10^8$ collisions (a). 
  \label{figRobAcc}
}
\end{figure}

Again we expect the exponential acceleration, as confirmed in Fig.
\ref{figRobAcc}. Note a relatively slow transition to $\beta=1$, which
could be easily missed if the simulation would be terminated after $10^7$
collisions.

\subsection{Annular billiard}

\begin{figure}
  \vspace{0.2cm}
  \includegraphics[width=4cm]{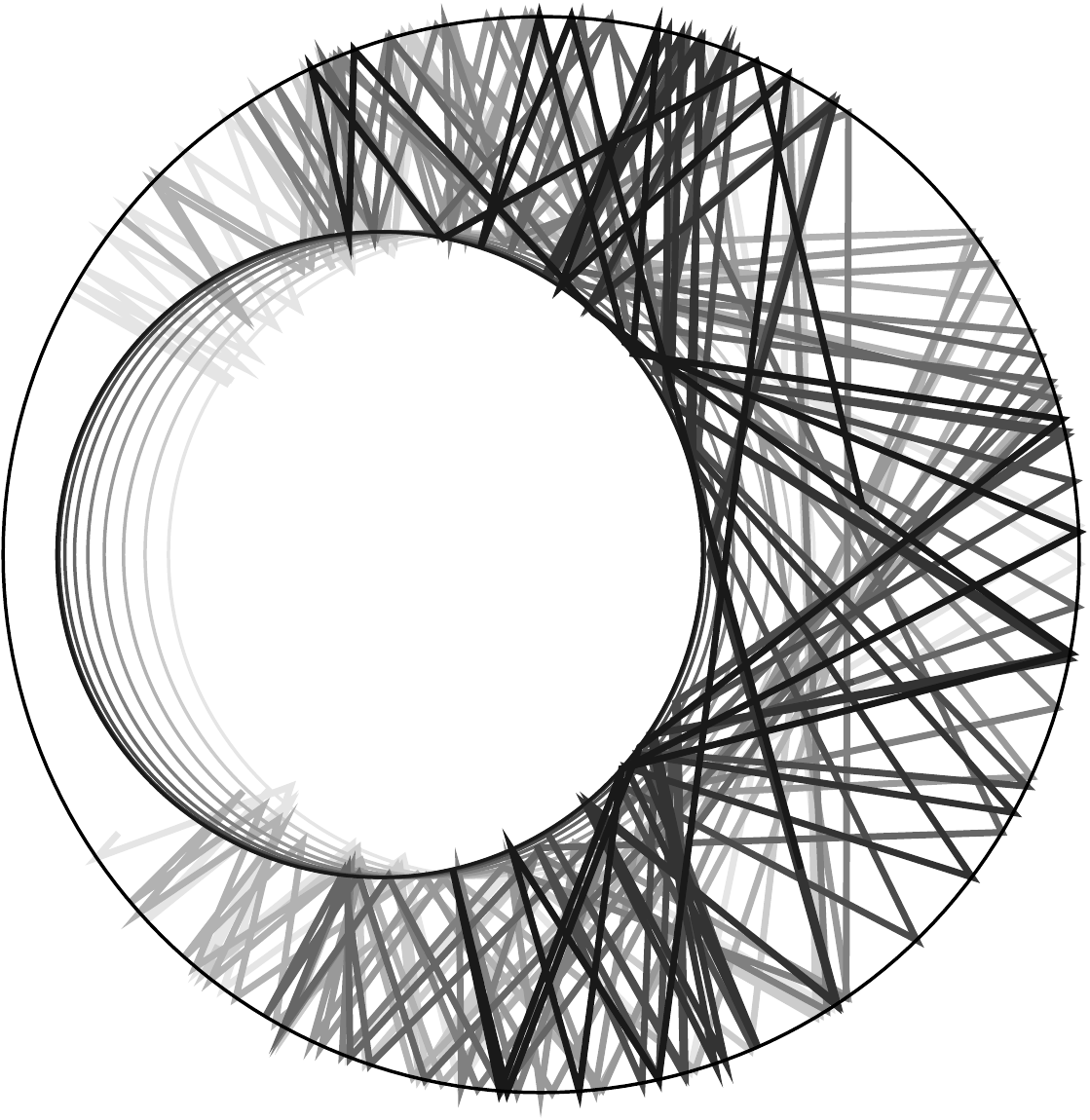}\hspace{0.5cm}
\includegraphics[height=4cm]{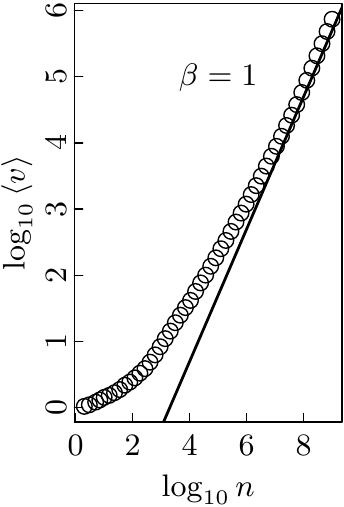}
\caption{
  A trajectory in the time dependent annular billiard
  and the evident exponential acceleration.
}
\label{figAnnBill}
\end{figure}

In this subsection we consider a time-dependent billiard which was
already studied in \cite{Carvalho2006b,Carvalho2011}. 
The billiard domain of the annular billiard is the interior of the circle
\begin{equation}
  x^2 + y^2 = R^2,
  \label{eqAnnCircR}
\end{equation}
with the circular hole 
\begin{equation}
  (x-d)^2 + y^2 = r^2,\quad r+d<R,
  \label{eqAnnDr}
\end{equation}
which is shifted from the center of the big circle by $d$. An example is
in Fig. \ref{figAnnBill}. Trajectories which do not hit the internal
boundary are regular as in a circle billiard, while the phase-space of
trajectories which hit the internal boundary is partially chaotic and
partially regular, depending on
the parameters $d$ and $r$.

In \cite{Carvalho2006b,Carvalho2011} authors studied a time-depend
annular billiard where
$d=0.5$ is fixed while $R=1+\epsilon\,\cos{t}$ and
$r=0.3+\epsilon\,\cos{t}$ depend on time, where $\epsilon=0.05$.
This particular example is almost completely chaotic, with relatively small
islands of regular motion, if the trajectories which do not hit the
internal boundary are excluded. Additionally, a relatively small
variations of the billiard (small deformation parameter $\epsilon$),
renders the mechanism of exponential acceleration even weaker. It is thus
no surprise that in  \cite{Carvalho2006b,Carvalho2011} 
the exponential acceleration was not observed even after
$10^8$ collisions, although a rather big acceleration
exponent $\beta\approx0.62$ was observed.

We have considered a different set of parameters for which a
phase-space of corresponding static billiards is more diverse. We took
fixed $R=1$ and $r=0.6$ and time-dependent $d=0.3\,\sin{t}$, Fig.
\ref{figAnnBill}. As expected, we observe a clear exponential acceleration,
Fig. \ref{figAnnBill}.

\section{Conclusions}

The central result of the paper is the following statement: If a phase
space structure of some corresponding static billiards of a generic time-dependent
billiard is of a mixed type, with
coexisting regular and chaotic domains, then, in the adiabatic regime,
such a time-dependent billiard exhibits exponential Fermi acceleration.
A nongeneric example, for which the exponential acceleration is not
possible, is a shape preserving time-dependent billiard \cite{Batistic2014}.
Since a phase-space structure of a typical
billiard is of a mixed type, we can conclude that the exponential acceleration is
a most common mode of acceleration in time-dependent billiards in
the adiabatic regime.

We have shown in this paper that in a time-dependent billiard a relevant part of the
dynamics of fast variables in the adiabatic regime 
can be represented as a stochastic hopping between
invariant components of corresponding static billiards where the hopping
probabilities are represented as a Markovian transition matrices.
The velocity dynamics is then described as an integral over a path through
the space of invariant components of corresponding static billiards.
We have shown that if a
number of possible paths through the space of invariant components grows
exponential with time, then this in general implies the exponential Fermi acceleration.
This should be typically observed in a mixed type billiards such as those
we have studied numerically in this paper.

Future studies should aim at a general understanding of 
transition probabilities between invariant components of corresponding
static billiards. These were
already calculated for a time-dependent mushroom billiard \cite{Gelfreich2013arxiv}. 
It is also important to understand a quantum-mechanical aspects of
time-dependent billiard in a semiclassical limit \cite{Schmelcher2009}, for which the formalism
presented in this paper could prove relevant.

\section*{Acknowledgement}

I would
like to thank Prof. Marko Robnik for useful discussions and careful reading
of the manuscript.
This work was supported by the Slovenian Research Agency ARRS. 

\section*{Appendix A}

Here we derive the field $f(\lega,\theta,t)$ introduced in
(\ref{eqPathInteg}).
First let us introduce a boundary function $h(\lega,t)$, which is
differentiable, zero on a billiard boundary, positive in a billiard domain
and negative outside a billiard domain.
Then 
\begin{equation}
    \mathbf{n}=\frac{\nabla h(\lega,t)}{\|\nabla h(\lega,t)\|}
    \label{eqNormala}
\end{equation}
is an inward normal unit vector at the point $(\lega,t)$ on the boundary.
A total derivative of $h(\lega,t)$ on a boundary is zero by
definition,
\begin{equation}
    \nabla h(\lega,t)\cdot \hitru+\parcia{h(\lega,t)}{t}=0,
    \label{eqTotalH}
\end{equation}
where $\hitru$ is the velocity of the boundary. 
From Eqs. (\ref{eqNormala}) and (\ref{eqTotalH}) we find a normal
component of a boundary velocity
\begin{equation}
    \mathbf{n}\cdot \hitru=-\frac{1}{\|\nabla
    h(\lega,t)\|}\,\parcia{h(\lega,t)}{t}.
    \label{eqTotalHposl}
\end{equation}
Let $\mathbf{k}=(\cos\theta,\sin\theta)$ be the unit vector in the
direction of the particle velocity. By the definition,
\begin{equation}
    \mathbf{k}=\frac{d \lega}{ds},
    \label{eqKdef}
\end{equation}
where $s$ is a geometrical length of the trajectory.
Now, according to Eq. (\ref{eqTestAprox}),
the change of the velocity at a collision can be written as
\begin{eqnarray}
    \Delta{v}&=&
    \left( \potem{\mathbf{k}}-\prej{\mathbf{k}}
    \right)\cdot\hitru\nonumber\\
    &=& 
    \left(\mathbf{n}\cdot\potem{\mathbf{k}}-\mathbf{n}\cdot\prej{\mathbf{k}}\right)\,\mathbf{n}\cdot\hitru\nonumber\\
    &=& 
    \left(|\mathbf{n}\cdot\potem{\mathbf{k}}|+|\mathbf{n}\cdot\prej{\mathbf{k}}|\right)\,\mathbf{n}\cdot\hitru
    \label{eqKolizon}
\end{eqnarray}
where the subscripts $-$ and $+$ denote a value right before and right
after a collision, respectively, and where we have taken into account that
$\mathbf{n}$ is always pointing inside the billiard domain. 

A change of the velocity appears only when the particle hits the
boundary, thus the field $f(\lega,t)$ can be defined in terms of the Dirac
delta function,
\begin{eqnarray}
    f(\lega,\theta,t)
    &=&
    \delta(h(\lega,\theta,t))\left|\frac{d\,h(\lega,t)}{ds}\right|\ldots\nonumber\\
    &=&
    \delta(h(\lega,\theta,t))\,
    \left|\nabla h(\lega,\theta,t)\cdot
    \mathbf{k}+\frac{1}{v}\parcia{h(\lega,t)}{t}\right|\ldots\nonumber\\
    &\approx& \delta{(h(\lega,t))}
    \left|\nabla h(\lega,\theta,t)\cdot \mathbf{k}\right|\ldots
    \label{eqFimaDelta}
\end{eqnarray}
where we have taken into account (\ref{eqKdef}) and $dt/ds=1/v$, and in
the last line neglect the term which is vanishing in the adiabatic limit.
Combining everything together gives finally
\begin{equation}
    f(\lega,\theta,t)=-2\,\delta(h(\lega,t))
    \,\frac{\left|\nabla h(\lega,t)\cdot\mathbf{k}\right|^2}{\|\nabla
    h(\lega,t)\|^2}\,\parcia{h(\lega,t)}{t}.
    \label{eqDefftocno}
\end{equation}
This is a possible form of the field $f(\lega,\theta,t)$, however, in
practice it is more convenient to work with a smooth field if we can
neglect the stepwise structure of the velocity dynamics.

\section*{Appendix B}

The velocity of the particle in a time-dependent billiard is a stepwise
function of time, with jumps at collisions of the particle with the
boundary. While the jumps are of the order of the velocity of the
boundary, the stepwise structure of the velocity dynamics becomes
unimportant in the adiabatic regime and can be as well represented with
some continuous curve.

We define a continuous velocity $v'$ of a trajectory as 
\begin{equation}
  v'(t)=\int_0^{t}dt\,v\,f'(\lega,\theta,t),
  \label{eqKompSmooth}
\end{equation}
where the field $f'(\lega,\theta,t)$ is defined in every phase-space point as
\begin{equation}
  f'(\lega,\theta,t)=-\frac{\lega_b-\lega_a}{\|\lega_b-\lega_a\|^2}\cdot(\hitru_b-\hitru_a)=-\dot{\ell}/\ell,
  \label{eqDefAccFact}
\end{equation}
where $\lega_b$ and $\lega_a$ are two intersections between the straight line passing
through $(\lega,\theta)$ and the boundary of the corresponding static billiard at
time $t$, while $\hitru_b$ and $\hitru_a$ are their velocities and
$\ell=\|\lega_b-\lega_a\|$ is their distance.
In other words, points $\lega_b$ and $\lega_a$ are two successive
collision points of a particle passing though $(\lega,\theta)$ in a
corresponding static billiard at time $t$.

That (\ref{eqKompSmooth}) approximates the true velocity can be
demonstrated as follows.
In every point $(\lega,\theta,t)$ on a trajectory between two successive collisions at
points $\lega_n$ and $\lega_{n-1}$,
we can approximate $\lega_b\approx\lega_n$ and $\lega_a\approx\lega_{n-1}$
up to corrections of the order of $1/v$,
from which it follows
\begin{eqnarray}
  v'_n-v'_{n-1} &=& \int_{t_{n-1}}^{t_{n}}
    dt\,v_{n-1}\,f'(\lega,\theta,t)\nonumber\\
    &\approx&
  -\frac{\lega_{n}-\lega_{n-1}}{\|\lega_{n}-\lega_{n-1}\|}\cdot(\hitru_{n}-\hitru_{n-1})
  \nonumber\\
  &=&-\frac{\hitr_{n-1}}{v_{n-1}}\cdot(\hitru_n-\hitru_{n-1}).
  \label{eqEvalIntf}
\end{eqnarray}
It is easy to see that the sum over the sequence of (\ref{eqTestAprox}) can be rearranged into
the sum over the sequence of (\ref{eqEvalIntf}), such that
\begin{equation}
    v_n-v_0\approx
    \frac{\hitr_{n}}{v_n}\cdot\hitru_{n} -\frac{\hitr_{0}}{v_0}\cdot\hitru_{0}
    +\int_{t_0}^{t_n} dt\,v\,f'(\lega,\theta,t).
    \label{eqSmoothV}
\end{equation}
Therefore, in the adiabatic limit, 
the continuous velocity $v'$ differs from the true
velocity $v$ by a
term proportional to the velocity of the boundary $\|\hitru\|$ plus an error from the adiabatic
approximation, which is vanishing.
Thus, neglecting the structures of the velocity dynamics on the resolution
$\|\hitru\|$, the velocity approximately satisfies the differential
equation
\begin{equation}
  \dot{v}=v\,f'(\lega,\theta,t),
  \label{eqKompFprime}
\end{equation}
and accordingly
\begin{equation}
  F=\int dt\,f'(\lega,\theta,t).
  \label{eqFprime}
\end{equation}

Consider now a shape-preserving time-dependent billiard in which
$f'(\lega,\theta,t)=f'(t)=\dot{\ell}/{\ell}$ depends only on time. 
If a driving is periodic, then $\ell$ is also periodic. Thus for every
possible trajectory
\begin{equation}
  F_1=\int_0^T dt\,f'(\lega,\theta,t)=\int_0^T dt\,f'(t)=0.
  \label{eqShapPresF1}
\end{equation}
Therefore, in a shape preserving time-dependent billiard the exponential
acceleration is not possible.

\bibliography{fermiacc.bib}

\begin{thebibliography}{38}%
\makeatletter
\providecommand \@ifxundefined [1]{%
 \@ifx{#1\undefined}
}%
\providecommand \@ifnum [1]{%
 \ifnum #1\expandafter \@firstoftwo
 \else \expandafter \@secondoftwo
 \fi
}%
\providecommand \@ifx [1]{%
 \ifx #1\expandafter \@firstoftwo
 \else \expandafter \@secondoftwo
 \fi
}%
\providecommand \natexlab [1]{#1}%
\providecommand \enquote  [1]{``#1''}%
\providecommand \bibnamefont  [1]{#1}%
\providecommand \bibfnamefont [1]{#1}%
\providecommand \citenamefont [1]{#1}%
\providecommand \href@noop [0]{\@secondoftwo}%
\providecommand \href [0]{\begingroup \@sanitize@url \@href}%
\providecommand \@href[1]{\@@startlink{#1}\@@href}%
\providecommand \@@href[1]{\endgroup#1\@@endlink}%
\providecommand \@sanitize@url [0]{\catcode `\\12\catcode `\$12\catcode
  `\&12\catcode `\#12\catcode `\^12\catcode `\_12\catcode `\%12\relax}%
\providecommand \@@startlink[1]{}%
\providecommand \@@endlink[0]{}%
\providecommand \url  [0]{\begingroup\@sanitize@url \@url }%
\providecommand \@url [1]{\endgroup\@href {#1}{\urlprefix }}%
\providecommand \urlprefix  [0]{URL }%
\providecommand \Eprint [0]{\href }%
\providecommand \doibase [0]{http://dx.doi.org/}%
\providecommand \selectlanguage [0]{\@gobble}%
\providecommand \bibinfo  [0]{\@secondoftwo}%
\providecommand \bibfield  [0]{\@secondoftwo}%
\providecommand \translation [1]{[#1]}%
\providecommand \BibitemOpen [0]{}%
\providecommand \bibitemStop [0]{}%
\providecommand \bibitemNoStop [0]{.\EOS\space}%
\providecommand \EOS [0]{\spacefactor3000\relax}%
\providecommand \BibitemShut  [1]{\csname bibitem#1\endcsname}%
\let\auto@bib@innerbib\@empty
\bibitem [{\citenamefont {Fermi}(1949)}]{Fermi1949}%
  \BibitemOpen
  \bibfield  {author} {\bibinfo {author} {\bibfnamefont {E.}~\bibnamefont
  {Fermi}},\ }\href {\doibase 10.1103/PhysRev.75.1169} {\bibfield  {journal}
  {\bibinfo  {journal} {Phys. Rev.}\ }\textbf {\bibinfo {volume} {75}},\
  \bibinfo {pages} {1169} (\bibinfo {year} {1949})}\BibitemShut {NoStop}%
\bibitem [{\citenamefont {Kotera}\ and\ \citenamefont
  {Olinto}(2011)}]{Kotera2011}%
  \BibitemOpen
  \bibfield  {author} {\bibinfo {author} {\bibfnamefont {K.}~\bibnamefont
  {Kotera}}\ and\ \bibinfo {author} {\bibfnamefont {A.~V.}\ \bibnamefont
  {Olinto}},\ }\href@noop {} {\bibfield  {journal} {\bibinfo  {journal} {Annual
  Review of Astronomy and Astrophysics}\ }\textbf {\bibinfo {volume} {49}},\
  \bibinfo {pages} {119} (\bibinfo {year} {2011})}\BibitemShut {NoStop}%
\bibitem [{\citenamefont {Bian}\ and\ \citenamefont {Kontar}(2013)}]{Bian2013}%
  \BibitemOpen
  \bibfield  {author} {\bibinfo {author} {\bibfnamefont {N.~H.}\ \bibnamefont
  {Bian}}\ and\ \bibinfo {author} {\bibfnamefont {E.~P.}\ \bibnamefont
  {Kontar}},\ }\href {\doibase 10.1103/PhysRevLett.110.151101} {\bibfield
  {journal} {\bibinfo  {journal} {Phys. Rev. Lett.}\ }\textbf {\bibinfo
  {volume} {110}},\ \bibinfo {pages} {151101} (\bibinfo {year}
  {2013})}\BibitemShut {NoStop}%
\bibitem [{\citenamefont {Saif}\ \emph {et~al.}(1998)\citenamefont {Saif},
  \citenamefont {Bialynicki-Birula}, \citenamefont {Fortunato},\ and\
  \citenamefont {Schleich}}]{Saif1998}%
  \BibitemOpen
  \bibfield  {author} {\bibinfo {author} {\bibfnamefont {F.}~\bibnamefont
  {Saif}}, \bibinfo {author} {\bibfnamefont {I.}~\bibnamefont
  {Bialynicki-Birula}}, \bibinfo {author} {\bibfnamefont {M.}~\bibnamefont
  {Fortunato}}, \ and\ \bibinfo {author} {\bibfnamefont {W.~P.}\ \bibnamefont
  {Schleich}},\ }\href {\doibase 10.1103/PhysRevA.58.4779} {\bibfield
  {journal} {\bibinfo  {journal} {Phys. Rev. A}\ }\textbf {\bibinfo {volume}
  {58}},\ \bibinfo {pages} {4779} (\bibinfo {year} {1998})}\BibitemShut
  {NoStop}%
\bibitem [{\citenamefont {Tabachnikov}(2005)}]{Tabachnikov}%
  \BibitemOpen
  \bibfield  {author} {\bibinfo {author} {\bibfnamefont {S.}~\bibnamefont
  {Tabachnikov}},\ }\href@noop {} {\emph {\bibinfo {title} {Geometry and
  Billiards}}},\ \bibinfo {series} {Students Mathematical Library},
  Vol.~\bibinfo {volume} {30}\ (\bibinfo  {publisher} {AMS},\ \bibinfo {year}
  {2005})\BibitemShut {NoStop}%
\bibitem [{\citenamefont {Chernov}\ and\ \citenamefont
  {Markarian}(2006)}]{Chernov}%
  \BibitemOpen
  \bibfield  {author} {\bibinfo {author} {\bibfnamefont {N.}~\bibnamefont
  {Chernov}}\ and\ \bibinfo {author} {\bibfnamefont {R.}~\bibnamefont
  {Markarian}},\ }\href@noop {} {\emph {\bibinfo {title} {Chaotic
  Billiards}}},\ \bibinfo {series} {Mathematical Survey and Monographs}, Vol.\
  \bibinfo {volume} {127}\ (\bibinfo  {publisher} {AMS},\ \bibinfo {year}
  {2006})\BibitemShut {NoStop}%
\bibitem [{\citenamefont {Koiller}\ \emph {et~al.}(1995)\citenamefont
  {Koiller}, \citenamefont {Markarian}, \citenamefont {Oliffson~Kamphorst},\
  and\ \citenamefont {Pinto~de Carvalho}}]{Koiller1995}%
  \BibitemOpen
  \bibfield  {author} {\bibinfo {author} {\bibfnamefont {J.}~\bibnamefont
  {Koiller}}, \bibinfo {author} {\bibfnamefont {R.}~\bibnamefont {Markarian}},
  \bibinfo {author} {\bibfnamefont {S.}~\bibnamefont {Oliffson~Kamphorst}}, \
  and\ \bibinfo {author} {\bibfnamefont {S.}~\bibnamefont {Pinto~de
  Carvalho}},\ }\href {http://stacks.iop.org/0951-7715/8/983} {\bibfield
  {journal} {\bibinfo  {journal} {Nonlinearity}\ }\textbf {\bibinfo {volume}
  {8}},\ \bibinfo {pages} {983} (\bibinfo {year} {1995})}\BibitemShut {NoStop}%
\bibitem [{\citenamefont {Batisti\'{c}}\ and\ \citenamefont
  {Robnik}(2013)}]{Batistic2013}%
  \BibitemOpen
  \bibfield  {author} {\bibinfo {author} {\bibfnamefont {B.}~\bibnamefont
  {Batisti\'{c}}}\ and\ \bibinfo {author} {\bibfnamefont {M.}~\bibnamefont
  {Robnik}},\ }\href {\doibase 10.1103/PhysRevE.88.052913} {\bibfield
  {journal} {\bibinfo  {journal} {Phys. Rev. E}\ }\textbf {\bibinfo {volume}
  {88}},\ \bibinfo {pages} {052913} (\bibinfo {year} {2013})}\BibitemShut
  {NoStop}%
\bibitem [{\citenamefont {Liss}\ \emph {et~al.}(2013)\citenamefont {Liss},
  \citenamefont {Liebchen},\ and\ \citenamefont {Schmelcher}}]{Liss2013}%
  \BibitemOpen
  \bibfield  {author} {\bibinfo {author} {\bibfnamefont {J.}~\bibnamefont
  {Liss}}, \bibinfo {author} {\bibfnamefont {B.}~\bibnamefont {Liebchen}}, \
  and\ \bibinfo {author} {\bibfnamefont {P.}~\bibnamefont {Schmelcher}},\
  }\href {\doibase 10.1103/PhysRevE.87.012912} {\bibfield  {journal} {\bibinfo
  {journal} {Phys. Rev. E}\ }\textbf {\bibinfo {volume} {87}},\ \bibinfo
  {pages} {012912} (\bibinfo {year} {2013})}\BibitemShut {NoStop}%
\bibitem [{\citenamefont {St\"ockmann}(1999)}]{stoe}%
  \BibitemOpen
  \bibfield  {author} {\bibinfo {author} {\bibfnamefont {H.-J.}\ \bibnamefont
  {St\"ockmann}},\ }\href@noop {} {\emph {\bibinfo {title} {Quantum Chaos - An
  Introduction}}}\ (\bibinfo  {publisher} {Cambridge: Cambridge University
  Press},\ \bibinfo {year} {1999})\BibitemShut {NoStop}%
\bibitem [{\citenamefont {Gmachl}\ \emph {et~al.}(1998)\citenamefont {Gmachl},
  \citenamefont {Capasso}, \citenamefont {Narimanov}, \citenamefont {N\"ockel},
  \citenamefont {Stone}, \citenamefont {Faist}, \citenamefont {Sivco},\ and\
  \citenamefont {Cho}}]{Gmachl1998}%
  \BibitemOpen
  \bibfield  {author} {\bibinfo {author} {\bibfnamefont {C.}~\bibnamefont
  {Gmachl}}, \bibinfo {author} {\bibfnamefont {F.}~\bibnamefont {Capasso}},
  \bibinfo {author} {\bibfnamefont {E.~E.}\ \bibnamefont {Narimanov}}, \bibinfo
  {author} {\bibfnamefont {J.~U.}\ \bibnamefont {N\"ockel}}, \bibinfo {author}
  {\bibfnamefont {A.~D.}\ \bibnamefont {Stone}}, \bibinfo {author}
  {\bibfnamefont {J.}~\bibnamefont {Faist}}, \bibinfo {author} {\bibfnamefont
  {D.~L.}\ \bibnamefont {Sivco}}, \ and\ \bibinfo {author} {\bibfnamefont
  {A.~Y.}\ \bibnamefont {Cho}},\ }\href@noop {} {\bibfield  {journal} {\bibinfo
   {journal} {Science}\ }\textbf {\bibinfo {volume} {280}},\ \bibinfo {pages}
  {1556} (\bibinfo {year} {1998})}\BibitemShut {NoStop}%
\bibitem [{\citenamefont {Ponomarenko}\ \emph {et~al.}(2008)\citenamefont
  {Ponomarenko}, \citenamefont {Schedin}, \citenamefont {Katsnelson},
  \citenamefont {Yang}, \citenamefont {Hill}, \citenamefont {Novoselov},\ and\
  \citenamefont {Geim}}]{Ponomarenko2008}%
  \BibitemOpen
  \bibfield  {author} {\bibinfo {author} {\bibfnamefont {L.~A.}\ \bibnamefont
  {Ponomarenko}}, \bibinfo {author} {\bibfnamefont {F.}~\bibnamefont
  {Schedin}}, \bibinfo {author} {\bibfnamefont {M.~I.}\ \bibnamefont
  {Katsnelson}}, \bibinfo {author} {\bibfnamefont {R.}~\bibnamefont {Yang}},
  \bibinfo {author} {\bibfnamefont {E.~W.}\ \bibnamefont {Hill}}, \bibinfo
  {author} {\bibfnamefont {K.~S.}\ \bibnamefont {Novoselov}}, \ and\ \bibinfo
  {author} {\bibfnamefont {A.~K.}\ \bibnamefont {Geim}},\ }\href {\doibase
  10.1126/science.1154663} {\bibfield  {journal} {\bibinfo  {journal}
  {Science}\ }\textbf {\bibinfo {volume} {320}},\ \bibinfo {pages} {356}
  (\bibinfo {year} {2008})}\BibitemShut {NoStop}%
\bibitem [{\citenamefont {Ulam}(1961)}]{Ulam1961}%
  \BibitemOpen
  \bibfield  {author} {\bibinfo {author} {\bibfnamefont {S.~M.}\ \bibnamefont
  {Ulam}},\ }in\ \href@noop {} {\emph {\bibinfo {booktitle} {Proc. Fourth
  Berkeley Symp. on Math. Statist. and Prob.}}},\ Vol.~\bibinfo {volume} {3},\
  \bibinfo {editor} {edited by\ \bibinfo {editor} {\bibfnamefont
  {J.}~\bibnamefont {Neyman}}}\ (\bibinfo  {publisher} {Univ. of Calif.
  Press},\ \bibinfo {year} {1961})\ pp.\ \bibinfo {pages}
  {315--320}\BibitemShut {NoStop}%
\bibitem [{\citenamefont {Lieberman}\ and\ \citenamefont
  {Lichtenberg}(1972)}]{Lieberman1972}%
  \BibitemOpen
  \bibfield  {author} {\bibinfo {author} {\bibfnamefont {M.~A.}\ \bibnamefont
  {Lieberman}}\ and\ \bibinfo {author} {\bibfnamefont {A.~J.}\ \bibnamefont
  {Lichtenberg}},\ }\href {\doibase 10.1103/PhysRevA.5.1852} {\bibfield
  {journal} {\bibinfo  {journal} {Phys. Rev. A}\ }\textbf {\bibinfo {volume}
  {5}},\ \bibinfo {pages} {1852} (\bibinfo {year} {1972})}\BibitemShut
  {NoStop}%
\bibitem [{\citenamefont {Loskutov}\ \emph {et~al.}(1999)\citenamefont
  {Loskutov}, \citenamefont {Ryabov},\ and\ \citenamefont
  {Akinshin}}]{Loskutov1999}%
  \BibitemOpen
  \bibfield  {author} {\bibinfo {author} {\bibfnamefont {A.}~\bibnamefont
  {Loskutov}}, \bibinfo {author} {\bibfnamefont {A.}~\bibnamefont {Ryabov}}, \
  and\ \bibinfo {author} {\bibfnamefont {L.}~\bibnamefont {Akinshin}},\ }\href
  {\doibase 10.1134/1.558939} {\bibfield  {journal} {\bibinfo  {journal} {J.
  Exp. Theor. Phys.}\ }\textbf {\bibinfo {volume} {89}},\ \bibinfo {pages}
  {966} (\bibinfo {year} {1999})}\BibitemShut {NoStop}%
\bibitem [{\citenamefont {Lenz}\ \emph {et~al.}(2008)\citenamefont {Lenz},
  \citenamefont {Diakonos},\ and\ \citenamefont {Schmelcher}}]{Lenz2008}%
  \BibitemOpen
  \bibfield  {author} {\bibinfo {author} {\bibfnamefont {F.}~\bibnamefont
  {Lenz}}, \bibinfo {author} {\bibfnamefont {F.~K.}\ \bibnamefont {Diakonos}},
  \ and\ \bibinfo {author} {\bibfnamefont {P.}~\bibnamefont {Schmelcher}},\
  }\href {\doibase {10.1103/PhysRevLett.100.014103}} {\bibfield  {journal}
  {\bibinfo  {journal} {{Phys. Rev. Lett.}}\ }\textbf {\bibinfo {volume}
  {{100}}},\ \bibinfo {pages} {{014103}} (\bibinfo {year}
  {{2008}})}\BibitemShut {NoStop}%
\bibitem [{\citenamefont {Shah}(2011)}]{Shah2011}%
  \BibitemOpen
  \bibfield  {author} {\bibinfo {author} {\bibfnamefont {K.}~\bibnamefont
  {Shah}},\ }\href {\doibase {10.1103/PhysRevE.83.046215}} {\bibfield
  {journal} {\bibinfo  {journal} {{Phys. Rev. E}}\ }\textbf {\bibinfo {volume}
  {{83}}},\ \bibinfo {pages} {{046215}} (\bibinfo {year} {{2011}})}\BibitemShut
  {NoStop}%
\bibitem [{\citenamefont {Leonel}\ \emph {et~al.}(2009)\citenamefont {Leonel},
  \citenamefont {Oliveira},\ and\ \citenamefont {Loskutov}}]{Leonel2009a}%
  \BibitemOpen
  \bibfield  {author} {\bibinfo {author} {\bibfnamefont {E.~D.}\ \bibnamefont
  {Leonel}}, \bibinfo {author} {\bibfnamefont {D.~F.~M.}\ \bibnamefont
  {Oliveira}}, \ and\ \bibinfo {author} {\bibfnamefont {A.}~\bibnamefont
  {Loskutov}},\ }\href {\doibase {10.1063/1.3227740}} {\bibfield  {journal}
  {\bibinfo  {journal} {{Chaos}}\ }\textbf {\bibinfo {volume} {{19}}},\
  \bibinfo {pages} {{033142}} (\bibinfo {year} {{2009}})}\BibitemShut {NoStop}%
\bibitem [{\citenamefont {Kamphorst}\ \emph {et~al.}(2007)\citenamefont
  {Kamphorst}, \citenamefont {Leonel},\ and\ \citenamefont
  {da~Silva}}]{Kamphorst2007}%
  \BibitemOpen
  \bibfield  {author} {\bibinfo {author} {\bibfnamefont {S.~O.}\ \bibnamefont
  {Kamphorst}}, \bibinfo {author} {\bibfnamefont {E.~D.}\ \bibnamefont
  {Leonel}}, \ and\ \bibinfo {author} {\bibfnamefont {J.~K.~L.}\ \bibnamefont
  {da~Silva}},\ }\href {http://stacks.iop.org/1751-8121/40/i=37/a=F02}
  {\bibfield  {journal} {\bibinfo  {journal} {J. Phys. A: Math. Theor.}\
  }\textbf {\bibinfo {volume} {40}},\ \bibinfo {pages} {F887} (\bibinfo {year}
  {2007})}\BibitemShut {NoStop}%
\bibitem [{\citenamefont {de~Carvalho}\ \emph {et~al.}(2006)\citenamefont
  {de~Carvalho}, \citenamefont {Souza},\ and\ \citenamefont
  {Leonel}}]{Carvalho2006b}%
  \BibitemOpen
  \bibfield  {author} {\bibinfo {author} {\bibfnamefont {R.~E.}\ \bibnamefont
  {de~Carvalho}}, \bibinfo {author} {\bibfnamefont {F.~C.}\ \bibnamefont
  {Souza}}, \ and\ \bibinfo {author} {\bibfnamefont {E.~D.}\ \bibnamefont
  {Leonel}},\ }\href {\doibase 10.1103/PhysRevE.73.066229} {\bibfield
  {journal} {\bibinfo  {journal} {Phys. Rev. E}\ }\textbf {\bibinfo {volume}
  {73}},\ \bibinfo {pages} {066229} (\bibinfo {year} {2006})}\BibitemShut
  {NoStop}%
\bibitem [{\citenamefont {Gelfreich}\ and\ \citenamefont
  {Turaev}(2008)}]{Gelfreich2008a}%
  \BibitemOpen
  \bibfield  {author} {\bibinfo {author} {\bibfnamefont {V.}~\bibnamefont
  {Gelfreich}}\ and\ \bibinfo {author} {\bibfnamefont {D.}~\bibnamefont
  {Turaev}},\ }\href {\doibase 10.1088/1751-8113/41/21/212003} {\bibfield
  {journal} {\bibinfo  {journal} {J. Phys. A}\ }\textbf {\bibinfo {volume}
  {41}},\ \bibinfo {pages} {212003, 6} (\bibinfo {year} {2008})}\BibitemShut
  {NoStop}%
\bibitem [{\citenamefont {Batisti{\'c}}\ and\ \citenamefont
  {Robnik}(2011)}]{Batistic2011}%
  \BibitemOpen
  \bibfield  {author} {\bibinfo {author} {\bibfnamefont {B.}~\bibnamefont
  {Batisti{\'c}}}\ and\ \bibinfo {author} {\bibfnamefont {M.}~\bibnamefont
  {Robnik}},\ }\href {\doibase 10.1088/1751-8113/44/36/365101} {\bibfield
  {journal} {\bibinfo  {journal} {J. Phys. A}\ }\textbf {\bibinfo {volume}
  {44}},\ \bibinfo {pages} {365101, 21} (\bibinfo {year} {2011})}\BibitemShut
  {NoStop}%
\bibitem [{\citenamefont {Batisti\'c}\ and\ \citenamefont
  {Robnik}(2012)}]{Batistic2012}%
  \BibitemOpen
  \bibfield  {author} {\bibinfo {author} {\bibfnamefont {B.}~\bibnamefont
  {Batisti\'c}}\ and\ \bibinfo {author} {\bibfnamefont {M.}~\bibnamefont
  {Robnik}},\ }in\ \href@noop {} {\emph {\bibinfo {booktitle} {Let's face chaos
  through nonlinear dynamics}}},\ \bibinfo {series} {AIP Conference
  Proceedings}, Vol.\ \bibinfo {volume} {1468}\ (\bibinfo  {publisher} {AIP,
  Melville, New York},\ \bibinfo {year} {2012})\BibitemShut {NoStop}%
\bibitem [{\citenamefont {Batisti\'{c}}(2014)}]{Batistic2014}%
  \BibitemOpen
  \bibfield  {author} {\bibinfo {author} {\bibfnamefont {B.}~\bibnamefont
  {Batisti\'{c}}},\ }\href {\doibase 10.1103/PhysRevE.89.022912} {\bibfield
  {journal} {\bibinfo  {journal} {Phys. Rev. E}\ }\textbf {\bibinfo {volume}
  {89}},\ \bibinfo {pages} {022912} (\bibinfo {year} {2014})}\BibitemShut
  {NoStop}%
\bibitem [{\citenamefont {Shah}\ \emph {et~al.}(2010)\citenamefont {Shah},
  \citenamefont {Turaev},\ and\ \citenamefont {Rom-Kedar}}]{Shah2010}%
  \BibitemOpen
  \bibfield  {author} {\bibinfo {author} {\bibfnamefont {K.}~\bibnamefont
  {Shah}}, \bibinfo {author} {\bibfnamefont {D.}~\bibnamefont {Turaev}}, \ and\
  \bibinfo {author} {\bibfnamefont {V.}~\bibnamefont {Rom-Kedar}},\ }\href
  {\doibase 10.1103/PhysRevE.81.056205} {\bibfield  {journal} {\bibinfo
  {journal} {Phys. Rev. E}\ }\textbf {\bibinfo {volume} {81}},\ \bibinfo
  {pages} {056205} (\bibinfo {year} {2010})}\BibitemShut {NoStop}%
\bibitem [{\citenamefont {Liebchen}\ \emph {et~al.}(2011)\citenamefont
  {Liebchen}, \citenamefont {Buechner}, \citenamefont {Petri}, \citenamefont
  {Diakonos}, \citenamefont {Lenz},\ and\ \citenamefont
  {Schmelcher}}]{Liebchen2011}%
  \BibitemOpen
  \bibfield  {author} {\bibinfo {author} {\bibfnamefont {B.}~\bibnamefont
  {Liebchen}}, \bibinfo {author} {\bibfnamefont {R.}~\bibnamefont {Buechner}},
  \bibinfo {author} {\bibfnamefont {C.}~\bibnamefont {Petri}}, \bibinfo
  {author} {\bibfnamefont {F.~K.}\ \bibnamefont {Diakonos}}, \bibinfo {author}
  {\bibfnamefont {F.}~\bibnamefont {Lenz}}, \ and\ \bibinfo {author}
  {\bibfnamefont {P.}~\bibnamefont {Schmelcher}},\ }\href {\doibase
  {10.1088/1367-2630/13/9/093039}} {\bibfield  {journal} {\bibinfo  {journal}
  {{New J. Phys.}}\ }\textbf {\bibinfo {volume} {{13}}},\ \bibinfo {pages}
  {{093039}} (\bibinfo {year} {{2011}})}\BibitemShut {NoStop}%
\bibitem [{\citenamefont {Shah}(2013)}]{Shah2013}%
  \BibitemOpen
  \bibfield  {author} {\bibinfo {author} {\bibfnamefont {K.}~\bibnamefont
  {Shah}},\ }\href {\doibase 10.1103/PhysRevE.88.024902} {\bibfield  {journal}
  {\bibinfo  {journal} {Phys. Rev. E}\ }\textbf {\bibinfo {volume} {88}},\
  \bibinfo {pages} {024902} (\bibinfo {year} {2013})}\BibitemShut {NoStop}%
\bibitem [{\citenamefont {Gelfreich}\ \emph {et~al.}(2011)\citenamefont
  {Gelfreich}, \citenamefont {Rom-Kedar}, \citenamefont {Shah},\ and\
  \citenamefont {Turaev}}]{Gelfreich2011}%
  \BibitemOpen
  \bibfield  {author} {\bibinfo {author} {\bibfnamefont {V.}~\bibnamefont
  {Gelfreich}}, \bibinfo {author} {\bibfnamefont {V.}~\bibnamefont
  {Rom-Kedar}}, \bibinfo {author} {\bibfnamefont {K.}~\bibnamefont {Shah}}, \
  and\ \bibinfo {author} {\bibfnamefont {D.}~\bibnamefont {Turaev}},\ }\href
  {\doibase {10.1103/PhysRevLett.106.074101}} {\bibfield  {journal} {\bibinfo
  {journal} {{Phys. Rev. Lett.}}\ }\textbf {\bibinfo {volume} {{106}}},\
  \bibinfo {pages} {{074101}} (\bibinfo {year} {{2011}})}\BibitemShut {NoStop}%
\bibitem [{\citenamefont {Gelfreich}\ \emph {et~al.}(2012)\citenamefont
  {Gelfreich}, \citenamefont {Rom-Kedar},\ and\ \citenamefont
  {Turaev}}]{Gelfreich2012}%
  \BibitemOpen
  \bibfield  {author} {\bibinfo {author} {\bibfnamefont {V.}~\bibnamefont
  {Gelfreich}}, \bibinfo {author} {\bibfnamefont {V.}~\bibnamefont
  {Rom-Kedar}}, \ and\ \bibinfo {author} {\bibfnamefont {D.}~\bibnamefont
  {Turaev}},\ }\href
  {http://scitation.aip.org/content/aip/journal/chaos/22/3/10.1063/1.4736542}
  {\bibfield  {journal} {\bibinfo  {journal} {Chaos: An Interdisciplinary
  Journal of Nonlinear Science}\ }\textbf {\bibinfo {volume} {22}},\ \bibinfo
  {eid} {033116} (\bibinfo {year} {2012})}\BibitemShut {NoStop}%
\bibitem [{\citenamefont {Gelfreich}\ \emph {et~al.}(2013)\citenamefont
  {Gelfreich}, \citenamefont {Rom-Kedar},\ and\ \citenamefont
  {Turaev}}]{Gelfreich2013arxiv}%
  \BibitemOpen
  \bibfield  {author} {\bibinfo {author} {\bibfnamefont {V.}~\bibnamefont
  {Gelfreich}}, \bibinfo {author} {\bibfnamefont {V.}~\bibnamefont
  {Rom-Kedar}}, \ and\ \bibinfo {author} {\bibfnamefont {D.}~\bibnamefont
  {Turaev}},\ }\href@noop {} {\  (\bibinfo {year} {{2013}})},\ \bibinfo {note}
  {{arXiv:1305.2624}}\BibitemShut {NoStop}%
\bibitem [{\citenamefont {Hertz}(1910)}]{Hertz1910}%
  \BibitemOpen
  \bibfield  {author} {\bibinfo {author} {\bibfnamefont {P.}~\bibnamefont
  {Hertz}},\ }\href@noop {} {\bibfield  {journal} {\bibinfo  {journal} {Annalen
  der Physik}\ }\textbf {\bibinfo {volume} {338}},\ \bibinfo {pages} {225}
  (\bibinfo {year} {1910})}\BibitemShut {NoStop}%
\bibitem [{\citenamefont {Lenz}\ \emph {et~al.}(2010)\citenamefont {Lenz},
  \citenamefont {Petri}, \citenamefont {Diakonos},\ and\ \citenamefont
  {Schmelcher}}]{Lenz2010}%
  \BibitemOpen
  \bibfield  {author} {\bibinfo {author} {\bibfnamefont {F.}~\bibnamefont
  {Lenz}}, \bibinfo {author} {\bibfnamefont {C.}~\bibnamefont {Petri}},
  \bibinfo {author} {\bibfnamefont {F.~K.}\ \bibnamefont {Diakonos}}, \ and\
  \bibinfo {author} {\bibfnamefont {P.}~\bibnamefont {Schmelcher}},\ }\href
  {\doibase 10.1103/PhysRevE.82.016206} {\bibfield  {journal} {\bibinfo
  {journal} {Phys. Rev. E (3)}\ }\textbf {\bibinfo {volume} {82}},\ \bibinfo
  {pages} {016206} (\bibinfo {year} {2010})}\BibitemShut {NoStop}%
\bibitem [{\citenamefont {Arnold}(1989)}]{Arnold}%
  \BibitemOpen
  \bibfield  {author} {\bibinfo {author} {\bibfnamefont {V.~I.}\ \bibnamefont
  {Arnold}},\ }\href@noop {} {\emph {\bibinfo {title} {Mathematical Methods Of
  Classical Mechanics}}}\ (\bibinfo  {publisher} {Springer, New York},\
  \bibinfo {year} {1989})\BibitemShut {NoStop}%
\bibitem [{\citenamefont {Lenz}\ \emph {et~al.}(2009)\citenamefont {Lenz},
  \citenamefont {Petri}, \citenamefont {Koch}, \citenamefont {Diakonos},\ and\
  \citenamefont {Schmelcher}}]{Lenz2009}%
  \BibitemOpen
  \bibfield  {author} {\bibinfo {author} {\bibfnamefont {F.}~\bibnamefont
  {Lenz}}, \bibinfo {author} {\bibfnamefont {C.}~\bibnamefont {Petri}},
  \bibinfo {author} {\bibfnamefont {F.~R.~N.}\ \bibnamefont {Koch}}, \bibinfo
  {author} {\bibfnamefont {F.~K.}\ \bibnamefont {Diakonos}}, \ and\ \bibinfo
  {author} {\bibfnamefont {P.}~\bibnamefont {Schmelcher}},\ }\href@noop {}
  {\bibfield  {journal} {\bibinfo  {journal} {New J. Phys.}\ }\textbf {\bibinfo
  {volume} {11}},\ \bibinfo {pages} {083035} (\bibinfo {year}
  {2009})}\BibitemShut {NoStop}%
\bibitem [{\citenamefont {de~Carvalho}\ and\ \citenamefont
  {Abud}(2011)}]{Carvalho2011}%
  \BibitemOpen
  \bibfield  {author} {\bibinfo {author} {\bibfnamefont {R.~E.}\ \bibnamefont
  {de~Carvalho}}\ and\ \bibinfo {author} {\bibfnamefont {C.~V.}\ \bibnamefont
  {Abud}},\ }\href@noop {} {\bibfield  {journal} {\bibinfo  {journal} {Chaos
  Soliton. Fract.}\ }\textbf {\bibinfo {volume} {44}},\ \bibinfo {pages} {569}
  (\bibinfo {year} {2011})}\BibitemShut {NoStop}%
\bibitem [{\citenamefont {Robnik}(1983)}]{Robnik1983}%
  \BibitemOpen
  \bibfield  {author} {\bibinfo {author} {\bibfnamefont {M.}~\bibnamefont
  {Robnik}},\ }\href {http://stacks.iop.org/0305-4470/16/i=17/a=014} {\bibfield
   {journal} {\bibinfo  {journal} {J. Phys. A: Math. Gen.}\ }\textbf {\bibinfo
  {volume} {16}},\ \bibinfo {pages} {3971} (\bibinfo {year}
  {1983})}\BibitemShut {NoStop}%
\bibitem [{\citenamefont {Hayli}\ \emph {et~al.}(1987)\citenamefont {Hayli},
  \citenamefont {Dumont}, \citenamefont {Moulin-Ollagnier},\ and\ \citenamefont
  {Strelcyn}}]{Hayli1987}%
  \BibitemOpen
  \bibfield  {author} {\bibinfo {author} {\bibfnamefont {A.}~\bibnamefont
  {Hayli}}, \bibinfo {author} {\bibfnamefont {T.}~\bibnamefont {Dumont}},
  \bibinfo {author} {\bibfnamefont {J.}~\bibnamefont {Moulin-Ollagnier}}, \
  and\ \bibinfo {author} {\bibfnamefont {J.~M.}\ \bibnamefont {Strelcyn}},\
  }\href {http://stacks.iop.org/0305-4470/20/i=11/a=027} {\bibfield  {journal}
  {\bibinfo  {journal} {Journal of Physics A: Mathematical and General}\
  }\textbf {\bibinfo {volume} {20}},\ \bibinfo {pages} {3237} (\bibinfo {year}
  {1987})}\BibitemShut {NoStop}%
\bibitem [{\citenamefont {Schmelcher}\ \emph {et~al.}(2009)\citenamefont
  {Schmelcher}, \citenamefont {Lenz}, \citenamefont {Matrasulov}, \citenamefont
  {Sobirov},\ and\ \citenamefont {Avazbaev}}]{Schmelcher2009}%
  \BibitemOpen
  \bibfield  {author} {\bibinfo {author} {\bibfnamefont {P.}~\bibnamefont
  {Schmelcher}}, \bibinfo {author} {\bibfnamefont {F.}~\bibnamefont {Lenz}},
  \bibinfo {author} {\bibfnamefont {D.}~\bibnamefont {Matrasulov}}, \bibinfo
  {author} {\bibfnamefont {Z.}~\bibnamefont {Sobirov}}, \ and\ \bibinfo
  {author} {\bibfnamefont {S.}~\bibnamefont {Avazbaev}},\ }in\ \href {\doibase
  10.1007/978-90-481-3120-4_7} {\emph {\bibinfo {booktitle} {Complex Phenomena
  in Nanoscale Systems}}},\ \bibinfo {series and number} {NATO Science for
  Peace and Security Series B: Physics and Biophysics},\ \bibinfo {editor}
  {edited by\ \bibinfo {editor} {\bibfnamefont {G.}~\bibnamefont {Casati}}\
  and\ \bibinfo {editor} {\bibfnamefont {D.}~\bibnamefont {Matrasulov}}}\
  (\bibinfo  {publisher} {Springer Netherlands},\ \bibinfo {year} {2009})\ pp.\
  \bibinfo {pages} {81--95}\BibitemShut {NoStop}%
\end{thebibliography}%

\end{document}